\documentclass[journal]{IEEEtran}
\usepackage{cite}
\usepackage{amsmath,amssymb,amsfonts}
\usepackage{algorithmic}
\usepackage{graphicx}
\usepackage{textcomp}
\def\BibTeX{{\rm B\kern-.05em{\sc i\kern-.025em b}\kern-.08em
    T\kern-.1667em\lower.7ex\hbox{E}\kern-.125emX}}
\usepackage{tabularx}
\usepackage[utf8]{inputenc}
\usepackage{url}

\usepackage{listings}
\usepackage{color}
\usepackage[table]{xcolor}
\usepackage{transparent}
\usepackage{booktabs}
\usepackage{algorithm}

\usepackage{tikz}

\graphicspath{{./graphics/}}
\makeatletter
\def\input@path{{./graphics/}}
\makeatother

\usepackage{eqparbox}

\usepackage[inline]{enumitem}
\usepackage{subcaption}

\newcommand{\ie}{\emph{i.e.}}
\newcommand{\eg}{\emph{e.g.}}

\newcommand{\foi}{\emph{foi}}

\newenvironment{application}{\noindent\textbf{Application} \\ \begin{em}}{\end{em}}

\newcommand{\conv}{\textrm{cv}}
\newcommand{\dv}{\textrm{dv}}
\renewcommand{\path}{\mathbb{P}}
\newcommand{\subpath}{\mathbb{S}}

\newcommand{\ibg}{\mathcal{G}}
\newcommand{\IB}{\textit{IB}}
\newcommand{\DB}{\textit{DB}}

\newtheorem{Theorem}{Theorem}[section]

\newtheorem{Definition}{Definition}[section]

\newtheorem{Property}{Property}[section]

\definecolor{myviolet}{rgb}{0.8,0,1}
\definecolor{mygreen}{rgb}{0,0.8,0}
\definecolor{myorange}{rgb}{1,0.5,0}
\definecolor{lightgreen}{rgb}{0.9, 1.0, 0.9}
\definecolor{lightred}{rgb}{1.0, 0.9, 0.9}
\definecolor{darkgreen}{RGB}{0,120,0}
\newlength{\figurewidth}
\begin{document}
\renewcommand\labelitemi{$\bullet$}
\renewcommand{\arraystretch}{1.5}
\title{Graph-based Approach for Buffer-aware Timing Analysis of Heterogeneous Wormhole NoCs under Bursty Traffic}

\author{Fr\'ed\'eric Giroudot, Ahlem Mifdaoui}

\markboth{IEEEAccess}%
{something else here}


\maketitle

\begin{abstract}
This paper addresses the problem of worst-case timing analysis of heterogeneous wormhole NoCs, i.e., routers with different buffer sizes and transmission speeds, when consecutive-packet queuing (CPQ) occurs. The latter means that there are several consecutive packets of one flow queuing in the network. This scenario happens in the case of bursty traffic but also for non-schedulable traffic. Conducting such an analysis is known to be a challenging issue due to the sophisticated congestion patterns when enabling backpressure mechanisms. We tackle this problem through extending the applicability domain of our previous work for computing maximum delay bounds using Network Calculus, called Buffer-aware worst-case Timing Analysis (BATA). We propose a new Graph-based approach to improve the analysis of indirect blocking due to backpressure, while capturing the CPQ effect and keeping the information about dependencies between flows. Furthermore, the introduced approach improves the computation of indirect-blocking delay bounds in terms of complexity and ensures the safety of these bounds even for non-schedulable traffic. We provide further insights into the tightness and complexity issues of worst-case delay bounds yielded by the extended BATA with the Graph-based approach, denoted G-BATA. Our assessments show that the complexity has decreased by up to 100 times while offering an average tightness ratio of 71\%, with reference to the basic BATA. Finally, we evaluate the yielded improvements with G-BATA for a realistic use case against a recent state-of-the-art approach. This evaluation shows the applicability of G-BATA under more general assumptions and the impact of such a feature on the tightness and computation time. 
\end{abstract}


\section{Introduction}

Networks-on-chip (NoC) have become the standard interconnect for manycore architectures because of their high throughput and low latency capabilities. Most NoCs use wormhole routing \cite{survey_wormhole,mohapatra1998_wormhole} to transmit packets over the network: the packet is split in constant length words called \emph{flits}. Each flit is then forwarded from router to router, without having to wait for the remaining flits.
Compared to store and forward (S\&F) mechanisms, the wormhole routing allows to drastically reduce the storage buffers at each router \cite{kavaldjiev2003_survey_onchip_communications}, as well as the contention-free end-to-end delay of a packet, \ie{} almost insensitive to the packet path length. On the other hand, the wormhole routing complicates the possible congestion patterns, since a packet waiting for a resource to be freed can occupy several input buffers of routers along its path; thus introducing indirect blocking delays due to the \textit{buffer backpressure}\footnote{A logical mechanism to control the flow on a communication channel and avoid buffer overflow.} \cite{xiong_extending_rt_analysis_nocs}.

Hence, an appropriate timing analysis, taking into account these phenomena, has to be considered to provide safe delay bounds in wormhole NoCs.

Various timing analysis approaches of such NoCs have been proposed in the literature and a detailed qualitative benchmarking can be found in \cite{giroudot_buffer_aware}. The most relevant ones can broadly be categorized under three main classes: Scheduling Theory-based (\cite{burns_real_time_wormhole,burns_priority_share,liu_tighter_time_analysis,xiong_extending_rt_analysis_nocs, nikolic_rt_analysis_priority_preemptive_nocs}), Compositional Performance Analysis (CPA)-based (\cite{tobuschat_ernst_cpa_noc_backpressure,Rambo:2015:WCT}) and Network Calculus-based (\cite{qian_analysis_worst_case_delays_wormhole, jafari_ludb_vbr_flows_noc_vc,buffer_aware,boyer_computing_routes_kalray}).
However, these existing approaches suffer from some limitations, which are mainly due to: 
\begin{itemize}
\item \textit{considering specific assumptions}, such as: (i) distinct priorities and unique virtual channel assignment for each traffic flow in a router \cite{burns_real_time_wormhole} \cite{xiong_extending_rt_analysis_nocs}\cite{nikolic_rt_analysis_priority_preemptive_nocs}; (ii) a priority-share policy, but with a number of Virtual Channels (VC) at least equal to the number of traffic priority levels like in \cite{burns_priority_share} \cite{liu_tighter_time_analysis}\cite{qian_analysis_worst_case_delays_wormhole} \cite{tobuschat_ernst_cpa_noc_backpressure} or the maximum number of contentions along the NoC \cite{Nikolic:2013};
\item \textit{ignoring the buffer backpressure phenomena}, such as in \cite{Rambo:2015:WCT,jafari_ludb_vbr_flows_noc_vc,buffer_aware,burns_priority_share,kashif_2014_sla};
\item \textit{ignoring the flows serialization phenomena\footnote{The pipelined behavior of networks infers that the interference between flows along their shared subpaths should be counted only once, \ie{}, at their first convergence point.} along the flow path} by conducting an iterative response time computation, commonly used in Scheduling Theory and CPA, which generally leads to pessimistic delay bounds.
 \end{itemize}

Hence, to cope with these identified limitations, we have proposed in \cite{giroudot_buffer_aware} a timing analysis using Network Calculus \cite{le_boudec_thiran_nc_book} and referred as \textit{Buffer-Aware Worst-case Timing Analysis} (BATA) from this point on. The main idea of BATA consists in enhancing the delay bounds accuracy in wormhole NoCs through considering: (i) the flows serialization phenomena along the path of a \textit{flow of interest (\emph{foi})}, through considering the impact of interfering flows only at the first convergence point; (ii) refined interference patterns for the \emph{foi} accounting for the limited buffer size, through quantifying the way a packet can spread on a NoC with small buffers. Moreover, BATA is applicable for a large panel of wormhole NoCs: (i) routers implement a fixed priority arbitration of VCs; (ii) a VC can be assigned to an arbitrary number of traffic classes with different priority levels (\textit{VC sharing}); (iii) each traffic class may contain an arbitrary number of flows (\textit{priority sharing}). 

Nevertheless, this approach, along with many other state-of-the-art approaches in timing analysis of NoCs taking backpressure into account, considers only Constant Bit Rate (CBR) traffic, \ie{} one fixed-length packet within a minimum inter-arrival time. However, there are some traffic types, such as real-time audio, video and bursty data streams, which do not fulfill the CBR model. With such traffic, there can be more than one packet of the same flow consecutively queueing in the network. 
This scenario is referred to hereafter as consecutive-packet-queueing (CPQ). 
Assuming CPQ allows to consider bursty traffic flows, \ie{} flows that can inject several consecutive packets in the NoC, but also to cover scenarios where the network load is high or the traffic is non-schedulable so that a packet of one flow is delayed enough to impact the next injected packet of the same flow. 
The impact of CPQ assumption on the interference patterns has been revealed in \cite{giroudot2018_wip_extending_bata}. 
Moreover, further insights into the computation issues of the worst-case delay bounds yielded by BATA have been provided in \cite{giroudot2019_tightness_and_computation_assessment}. 
The results reveal that BATA provides good delay bounds for medium-scale configurations within less than one hour, but its complexity increases dramatically for large-scale configurations.

In this paper, we extend the applicability domain of BATA to ensure that the computed delay bounds remain safe without any assumption on CPQ, in addition to considering heterogeneous NoCs, \ie{} buffer sizes, link capacities and processing delays may differ from one router to another. Furthermore, we cope with the complexity issue of BATA to enable the timing analysis of large-scale configurations within a more reasonable time.

\textbf{Contributions:} we introduce a new Graph-based approach to improve the analysis of indirect blocking due to backpressure, while capturing the CPQ effect, for heterogeneous NoCs. 
This introduced approach, denoted G-BATA for Graph-based Buffer-Aware Timing Analysis, decreases in addition the complexity of the timing analysis process. 
Furthermore, we provide deeper insights into the tightness and complexity issues of worst-case delay bounds yielded by G-BATA, when varying different system parameters. Our assessments show that the computation times with G-BATA are 10 to 100 times lower than with BATA. 
Moreover, the average measured tightness ratio (achieved worst-case delay using simulation over analytical worst-case delay bound) of G-BATA is 71\%. Finally, we evaluate the yielded improvements with G-BATA for an automotive use-case against a recent state-of-the-art approach. 
This evaluation shows the applicability of G-BATA under more general assumptions than the state-of-the-art approach and the impact of such a feature on the tightness and computation time.

The remainder of this paper is organized as follows. We first present the problem statement in Section \ref{sec:problem_statement}. 
Then, we detail the system model and some preliminaries in Section \ref{sec:system_model}. 
Section \ref{sec:interference_graph_approach} details our new approach, G-BATA, to handle heterogeneous NoCs and the impact of backpressure under CPQ assumption. Finally, we evaluate the complexity and tightness of our approach in Section \ref{sec:performance_eval}, and the yielded improvements with G-BATA for a realistic use case against a recent state-of-the-art approach in Section \ref{sec:case_study}.


\section{Problem Statement}
\label{sec:problem_statement}

\subsection{Illustrative Example of CPQ Effect}
\label{subsec:illustrative_example}

\begin{figure}[ht]
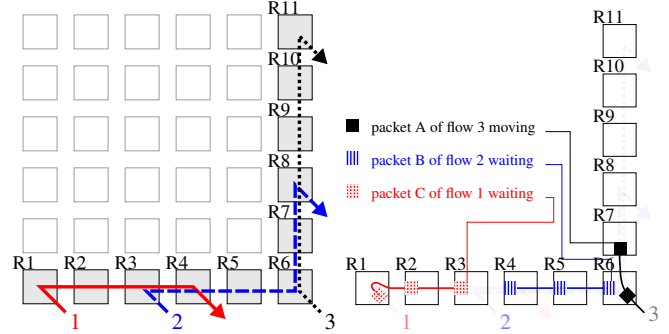

    \centering
    \resizebox{0.5\columnwidth}{!}{\input{WIP_intuitive_example.pdf_tex}}\resizebox{0.5\columnwidth}{!}{\input{WIP_intuitive_example_packet_detail.pdf_tex}}
    \caption{Example configuration (left) and packet stalling (right)}
    \label{fig:intuitive_example_config}
\end{figure}

The key element to take into account the backpressure phenomenon induced by limited buffer size is based on how packets can spread in the network when stalled.  
We consider an illlustrative example to better understand the impact of the buffer size on the packet spreading, and consequently the indirect blocking set (Figure \ref{fig:intuitive_example_config}).  
We make the following assumptions: \begin{enumerate*}[label=(\roman*)]\item each buffer can store only one flit; \item all flows have 3-flit-long packets; \item all flows are mapped to the same VC; \item the \emph{foi} is flow 1.\end{enumerate*}

We assume there is a packet A of flow 3 that has just been injected into the NoC and granted the use of the North output port of R6. 
Simultaneously, a packet B of flow 2 is reauesting the same output, but as A is already using it, B has to wait. 
B is stored in input buffers of R6, R5 and R4.
Finally, a packet C of flow 1 has reached R3 and now requests output port East of R3. However, the West input buffer of R4 is occupied by the tail flit of B.
Hence, C has to wait.
In that case, A indirectly blocks C, which means flow 3 can impact the transmission of flow 1 even though they do not share resources.
\begin{figure}[ht]
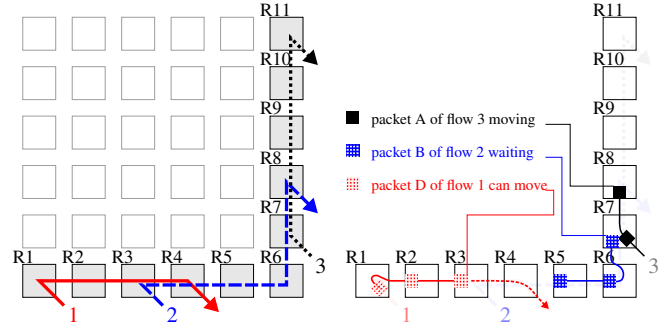

    \centering
    \resizebox{0.5\columnwidth}{!}{\input{WIP_problematic_example.pdf_tex}}\resizebox{0.5\columnwidth}{!}{\input{WIP_ib_example_single_packet_detail.pdf_tex}}
    \caption{Another configuration (left) where flow 1 cannot be blocked by flow 3 (right)}
    \label{fig:ib_example_single_packet}
\end{figure}

Now suppose flow 3 source is one hop further (Figure \ref{fig:ib_example_single_packet}). 
Consider that flow 3 has a packet A that has just been injected in the network and is using output port North of R7. As before, flow 2 has a packet B in the network that competes with A and has to wait. 
This time, however, the output requested by B is one hop further on the path of flow 2. 
As a result, B is stored in input buffers of R7, R6 and R5.
Finally, flow 1 has injected a packet C into the NoC.
Since B is stalling one hop further than before on its path, C can request ouput port East of R3 and use input buffer West of R4, and reach its destination without contention.

An approach that does not consider buffer sizes would predict that flow 3 could impact flow 1 regardless of the configuration. 
However, on the second example, we just showed such an assumption was pessimistic and could be avoided by taking buffer size into account. 
This illustrates the impact of the buffer size on packet spreading, and how buffer size reduces the section of the path on which a blocked packet can in its turn block another one. 

\begin{figure}[ht]
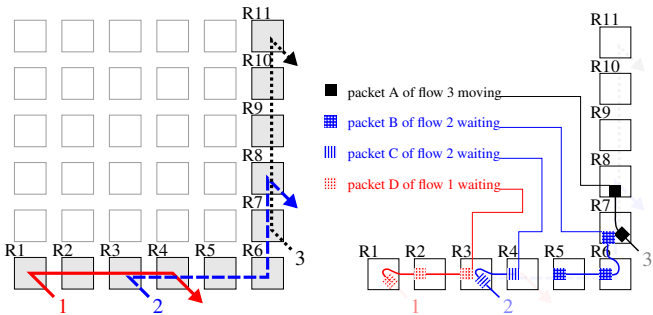

     \centering
     \resizebox{0.48\columnwidth}{!}{\input{WIP_problematic_example.pdf_tex}}\resizebox{0.52\columnwidth}{!}{\input{WIP_problematic_example_packet_detail.pdf_tex}}
     \caption{Example configuration (left) and CPQ not taken into account by BATA approach (right)}
     \label{fig:problematic_example_cpq}
\end{figure}

Still considering the same example, recalled on Figure \ref{fig:problematic_example_cpq}, we notice our analysis assumes there can only be one packet of flow 2 stalling in the network. 
Should there be an additional packet of flow 2 queueing right after the first one, the analysis would be different. 
We call such a scenario ``consecutive packet queueing'' (CPQ). 

To see how this limits the applicability of BATA approach, consider the packet configuration shown on Figure \ref{fig:problematic_example_cpq}. 
As before, a packet A of flow 3 is being transmitted. It requested and was granted output port North of R7. 
Flow 2 has a packet B also requesting output port North of R7 but as A is using it, it has to wait. 
B is stored in buffers in R5, R6 and R7. 
Moreover, there is an additional packet of flow 2, C, right behind B.
It was granted the use of output port East of R3 and is waiting at R4 for the next input buffer of its path to be available.
Finally, flow 1 has a packet D requesting the output port East of R3. Since C is already using this output, D has to wait.

Thus, packet D has to wait that packet A releases output port North of R7 to be able to move. 
This means flow 3 can indirectly block flow 1, even though BATA approach did not cover such a scenario. 
\bigskip

CPQ can happen when considering bursty traffic, \ie{} flows that can generate and inject a burst of several packets one after the other. Example of such flows include real-time audio and video streams. 
It can also occur when a packet of a periodic CBR flow experiences enough congestion for the next packet to ``catch up'' on it. 

\subsection{Identified Extensions of BATA}
To cover the CPQ assumption, we introduce the new Interference Graph approach, G-BATA (Graph-based BATA), which extends BATA with the following features:
\begin{itemize}
\item Generic system model to cover more general traffic pattern and heterogeneous NoC architectures;
\item Improved analysis of the backpressure phenomenon through refining the indirect blocking set computation;
\item Indirect blocking latency analysis taking into account the refined indirect blocking set.
\end{itemize}

Each identified extension will be detailed in the following sections and there will be illustrated through an example.

\section{Preliminaries and System Model}
\label{sec:system_model}

%
In this section, we detail the considered system model based on Network Calculus. First, we present the main concepts of Network Calculus that are used in this paper. Afterwards, we describe the network and flow models. Finally, we introduce the main definitions to cover the characteristics of heterogeneous NoCs with wormhole routing.
The notations will be introduced as they are needed and are also gathered in Table \ref{tab:notations}. As a general rule, upper indexes of a notation $X$ refer to a node or a subset of nodes, while lower indexes refer to a flow. $X_f^r$ means ``$X$ at node $r$ for flow $f$''.
\begin{table}[ht]
\centering
\begin{tabular}{@{}ll@{}}
\toprule
Notation         & Definition \\
\midrule
$\mathcal{F}$    & The set of flows on the NoC \\
$S_{flit}$       & The size of one flit        \\
$B^r$            & The buffer size at node $r$ \\
$\path_f$        & The list of nodes crossed by $f$ from source to destination \\
$\path_f[k]$     & The $k+1^{th}$ node of $f$ path         \\
$subpath(\path_k,\path_l)$   & The subpath of flow $k$ relatively to flow $l$ after $\dv(\path_k,\path_l)$ \\
$Last(\path_k,\path_l)$      & The index of $\dv(\path_k,\path_l)$ in $\path_k$ \\
$\conv(\path_k,\path_l)$     & The convergence node of $\path_k$ and $\path_l$ \\
$\dv(\path_k,\path_l)$       & The divergence node of $\path_k$ and $\path_l$\\
$f \ni r$        & Flow $f$ crosses node $r$ \\
$F \supset r$    & There is a flow $f \in F$ such that $f \ni r$ \\
$\mathbf{1}_{\{cdt\}}$ & equals $1$ if $cdt$ is true and zero otherwise \\
$L_f$            & The maximal packet length of $f$ \\
$J_f$            & The release jitter of $f$ \\
$P_f$            & The period of $f$ \\
$b_f$            & The number of packets in a burst of flow $f$ \\
$\alpha_f(t)$    & The initial arrival curve of $f$       \\
$ \alpha_f^r(t)$ & The arrival curve of $f$ at the input of node $r$   \\
$\sigma_f^r$     & The burst of $\alpha_f^r$   \\
$\rho_f^r$       & The rate of $\alpha_f^r$    \\
$ \beta_{f}(t)$  & The end-to-end service curve of $f$    \\ 
$\widetilde{\beta}_f^{subP}(t)$ & The VC-service curve of $f$ on $subP$ \\
$\widetilde{R}_f^{subP}$        & The rate of $\widetilde{\beta}_k^{subP} $   \\
$\widetilde{T}_f^{subP}$        & The latency of $\widetilde{\beta}_k^{subP} $   \\
$DB_f$           & The set of all flows directly interfering with $f$ \\
$DB_f^{|path}$   & Flows $i \in DB_f$ such that $\path_i \cap path \neq \varnothing$ \\
$hp(f)$          & Flows mapped to a VC of strict higher priority than $f$ \\
$sp(f)$          & Flows mapped to the same VC as $f$ \\
$lp(f)$          & Flows mapped to a VC of strict lower priority than $f$ \\
$slp(f)$         & All flows with a priority lower or equal than $f$ \\
$shp(f)$         & All flows with a priority higher or equal than $f$ \\
$IB_f$           & Indirect blocking set of flow $f$\\
$N_f^i$          & Nb of buffers to store a packet of $f$ from node $i$ on $\path_f$ \\
$D_f^{\path_f}$  & End-to-end delay bound of $f$ on $\path_f$ \\
\bottomrule
\end{tabular}
\smallskip
\caption{Summary of notations}
\label{tab:notations}
\end{table}

\subsection{Network Calculus Background}
\label{sec:network_calculus}

Network Calculus describes data flows by means of cumulative functions, defined as the number of transmitted bits during the time interval $[0, t]$.
Consider a system $S$ receiving input data flow with a Cumulative Arrival Function (CAF), $A(t)$, and putting out the same data flow with a Cumulative Departure Function (CDF), $D(t)$. 
To compute upper bounds on the worst-case delay and backlog, we need to introduce the maximum arrival curve, which provides an upper bound on the number of events, e.g., bits or packets, observed during any interval of time.
\begin{Definition}
\label{AC-def}
(Arrival Curve)\cite{le_boudec_thiran_nc_book} A function $\alpha$ is an arrival curve for a data flow with the CAF $A$, iff:
\[ \forall t,s \geq 0, s\leq t, A(t) - A(s) \leq \alpha(t-s) \]
\end{Definition}

A widely used curve is the leaky bucket curve, which guarantees a maximum burst $\sigma$ and a maximum rate $\rho$, \ie{}, the traffic flow is $(\sigma,\rho)$-constrained. In this case, the arrival curve is defined as $\gamma_{\sigma, \rho} (t)= \sigma+ \rho\cdot t$ for $t >0$.
Furthermore, we need to guarantee a minimum offered service within crossed nodes through the concept of minimum service curve.
\begin{Definition}
\label{SC-def}
(Simple Minimum Service Curve)\cite{le_boudec_thiran_nc_book} The function $\beta$ is the simple service curve for a data flow with the CAF $A$ and the CDF $D$, iff:
\[ \forall t \geq 0,D(t) \geq \inf_{s \leq t} (A(s)+ \beta(t-s)) \]
\end{Definition}

To define the leftover service curve for a flow crossing a node implementing aggregate scheduling, we need strict service curve property:
\begin{Definition}
\label{def:strict-service-curve}
(Strict service curve)\cite{le_boudec_thiran_nc_book} The function $\beta $ is a strict service curve for a data flow with the CDF $D(t)$, if for any backlogged period \footnote{A backlogged period $]s,t]$ is an interval of time during which the backlog is non null, \ie{}, $A(s)=D(s)$ and $\forall u \in ]s,t]$, $A(u)-D(u) > 0$} $]s,t]$, $D(t) - D(s) \geq \beta(t-s)$.
\end{Definition}
%

Knowing the arrival and service curves, one can compute the upper bounds on performance metrics for a data flow, according to the following theorem.
\begin{Theorem} (Performance Bounds)
\label{TH1}
Consider a flow constrained by an arrival curve $\alpha$ crossing a system $\mathcal{S}$ that offers a service curve $\beta$, then:\\
 Delay \footnote{$h(f,g)$: the maximum horizontal distance between $f$ and $g$}: $ \forall~t:~d(t)\leq h(\alpha,\beta)$ \\
Backlog \footnote{$v(f,g)$: the maximum vertical distance between $f$ and $g$}: $ \forall~t:~q(t)\leq v(\alpha,\beta)$ \\
Output arrival curve \footnote{$f \oslash g(t) = \sup_{\forall u \geq 0} \{ f(t+u) - g(u)\}$}: $\alpha^*(t) =\alpha \oslash\beta (t)$
\end{Theorem}

In the case of a leaky bucket arrival curve and a rate-latency service curve, the calculus of these bounds is greatly simplified. The delay and backlog are bounded by $\frac{\sigma}{R} + T$ and $ \sigma + \rho \cdot T$, respectively; and the output arrival curve is $\sigma+ \rho \cdot(T + t)$.

Finally, we need the following results concerning the end-to-end service curve of a flow of interest (\emph{foi}) accounting for flows serialization effects in feed-forward networks, based on the Pay Multiplexing Only Once (PMOO) principle \cite{schmitt_pmoo}, under non-preemptive Fixed Priority (FP) multiplexing.

\begin{Theorem}
The service curve offered to a flow of interest $f$ along its path $\path_f$, in a network under non-preemptive FP multiplexing with strict service curve nodes of the rate-latency type $\beta_{R,T}$ and leaky bucket constrained arrival curves $\alpha_{\sigma, \rho}$, is a rate-latency curve, with a rate $R^{\mathbb{P}_f}$ and a latency $T^{\mathbb{P}_f}$, as follows :
\label{Th:pmoo_fp}
\begin{subequations}
\label{pmoo_fp}
\begin{align}
&  R^{\mathbb{P}_f} = \min \limits_{k \in \mathbb{P}_f} \left( R^{k} - \sum \limits_{i \ni k, i \in shp(f)}{\rho_i} \right) \label{eq:pmoo_fp_R}\\
& T^{\mathbb{P}_f} =  \sum\limits_{k \in \mathbb{P}_f}  \left( T^{k} + \frac{\max\limits_{i \ni k, i \in slp(f)} L_i }{R^k}  \right) \nonumber \\
& + \sum\limits_{i \in DB_{f} \cap shp(f)} \frac{\sigma_i^{\conv(\path_i,\path_f)} + \rho_i \cdot \sum\limits_{k \in\mathbb{P}_f\cap \mathbb{P}_i} \left( T^{k} + \frac{\max\limits_{i \ni k, i \in slp(f)} L_i }{R^k} \right)    } {R^{\mathbb{P}_f}}  \label{eq:pmoo_fp_T}
\end{align}
\end{subequations}
where the required notations are defined in Table \ref{tab:notations}.
\end{Theorem}

\subsection{Network Model}
\label{subsec:network_model}
Our model can apply to an arbitrary NoC topology as long as the flows are routed in a deterministic, deadlock-free way (see \cite{survey_wormhole}), and in such a way that flows that interfere on their path do not interfere again after they diverge.
Nonetheless, we consider the commonly used 2D-mesh topology with input-buffered routers and XY-routing, known for their simplicity and high scalability.
Besides, XY-routing is widely used in COTS architectures (\eg{} \cite{wentzlaff_on_chip_interconnection_architecture_tile_processor}).

\begin{figure}[ht]
\centering
\resizebox{0.6\columnwidth}{!}{
\begingroup%
  \makeatletter%
  \providecommand\color[2][]{%
    \errmessage{(Inkscape) Color is used for the text in Inkscape, but the package 'color.sty' is not loaded}%
    \renewcommand\color[2][]{}%
  }%
  \providecommand\transparent[1]{%
    \errmessage{(Inkscape) Transparency is used (non-zero) for the text in Inkscape, but the package 'transparent.sty' is not loaded}%
    \renewcommand\transparent[1]{}%
  }%
  \providecommand\rotatebox[2]{#2}%
  \newcommand*\fsize{\dimexpr\f@size pt\relax}%
  \newcommand*\lineheight[1]{\fontsize{\fsize}{#1\fsize}\selectfont}%
  \ifx\svgwidth\undefined%
    \setlength{\unitlength}{184.71128869bp}%
    \ifx\svgscale\undefined%
      \relax%
    \else%
      \setlength{\unitlength}{\unitlength * \real{\svgscale}}%
    \fi%
  \else%
    \setlength{\unitlength}{\svgwidth}%
  \fi%
  \global\let\svgwidth\undefined%
  \global\let\svgscale\undefined%
  \makeatother%
  \begin{picture}(1,0.85135932)%
    \lineheight{1}%
    \setlength\tabcolsep{0pt}%
    \put(0,0){\includegraphics[width=\unitlength,page=1]{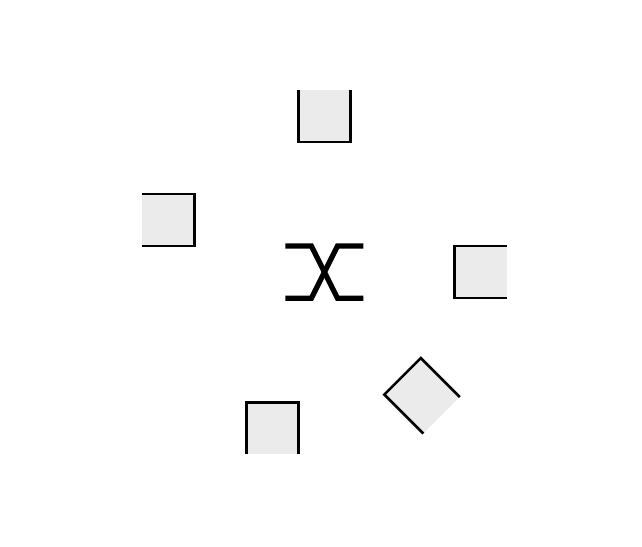}}%
    \put(0.40435354,0.0007187){\color[rgb]{0,0,0}\makebox(0,0)[lt]{\lineheight{1.25}\smash{\begin{tabular}[t]{l}South\end{tabular}}}}%
    \put(0,0){\includegraphics[width=\unitlength,page=2]{WIP_mesh_router_2D.pdf}}%
    \put(0.40435354,0.8127975){\color[rgb]{0,0,0}\makebox(0,0)[lt]{\lineheight{1.25}\smash{\begin{tabular}[t]{l}North\end{tabular}}}}%
    \put(0.89160066,0.42706042){\color[rgb]{0,0,0}\makebox(0,0)[lt]{\lineheight{1.25}\smash{\begin{tabular}[t]{l}East\end{tabular}}}}%
    \put(-0.00168522,0.42706042){\color[rgb]{0,0,0}\makebox(0,0)[lt]{\lineheight{1.25}\smash{\begin{tabular}[t]{l}West\end{tabular}}}}%
    \put(0.7900912,0.10222885){\color[rgb]{0,0,0}\makebox(0,0)[lt]{\lineheight{1.25}\smash{\begin{tabular}[t]{l}Local\end{tabular}}}}%
    \put(0,0){\includegraphics[width=\unitlength,page=3]{WIP_mesh_router_2D.pdf}}%
  \end{picture}%
\endgroup%
}
\caption{Typical 2D-mesh router}
\label{fig:2D_mesh_router}
\end{figure}

We consider typical input-buffered 2D-mesh routers with 5 pairs of input-output, namely North (N0, South (S), West (W), East (E) and Local (L), as shown on Figure \ref{fig:2D_mesh_router}.
Output-buffered routers have buffers located at the output ports instead of the input port but remain similar otherwise.

It is worth noticing that NoCs using output-buffered routers can be modeled similarly to input-buffered routers NoCs. The idea is that from a flow point of view, whether the buffer is located at the input or at the output does not change the number of buffers and links crossed by the flow on its path, as introduced in \cite{liu_buffer_aware}.

We also allow to model heterogeneous NoC architectures. For instance, we can specify distinct buffer sizes, link and router capacities and processing delays values on a single NoC. 

\begin{figure}
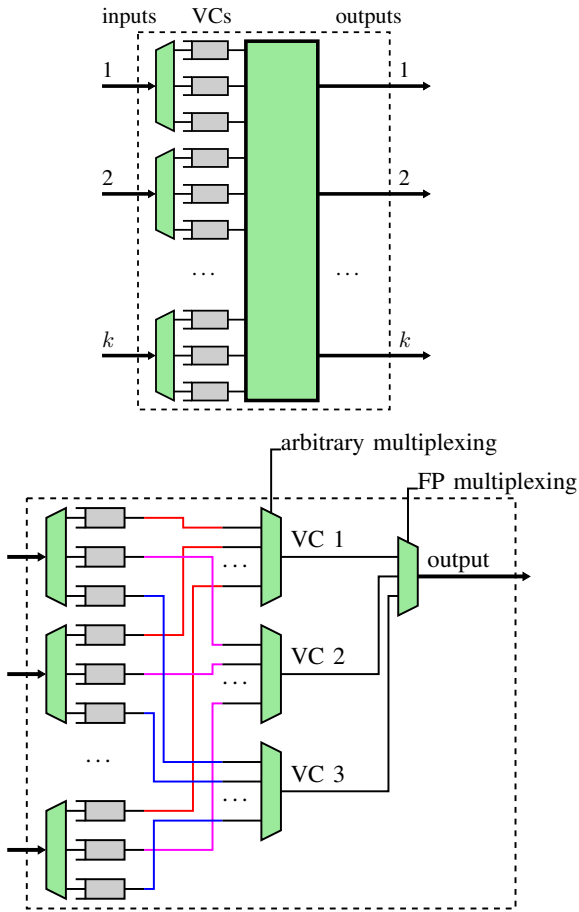

\centering
\resizebox{0.37\figurewidth}{!}{\input{NC_model_router.pdf_tex}}\\
\bigskip
\resizebox{0.58\figurewidth}{!}{\input{NC_model_output_multiplex.pdf_tex}}
\caption{Architecture of an input-buffered router (top) and output multiplexing (bottom) with the arbitration modeling choices}
\label{fig:router_arch}
\end{figure}

\begin{figure}
\centering
\rule{16pt}{0pt}\resizebox{\columnwidth}{!}{
\begingroup%
  \makeatletter%
  \providecommand\color[2][]{%
    \errmessage{(Inkscape) Color is used for the text in Inkscape, but the package 'color.sty' is not loaded}%
    \renewcommand\color[2][]{}%
  }%
  \providecommand\transparent[1]{%
    \errmessage{(Inkscape) Transparency is used (non-zero) for the text in Inkscape, but the package 'transparent.sty' is not loaded}%
    \renewcommand\transparent[1]{}%
  }%
  \providecommand\rotatebox[2]{#2}%
  \newcommand*\fsize{\dimexpr\f@size pt\relax}%
  \newcommand*\lineheight[1]{\fontsize{\fsize}{#1\fsize}\selectfont}%
  \ifx\svgwidth\undefined%
    \setlength{\unitlength}{324.0234375bp}%
    \ifx\svgscale\undefined%
      \relax%
    \else%
      \setlength{\unitlength}{\unitlength * \real{\svgscale}}%
    \fi%
  \else%
    \setlength{\unitlength}{\svgwidth}%
  \fi%
  \global\let\svgwidth\undefined%
  \global\let\svgscale\undefined%
  \makeatother%
  \begin{picture}(1,0.38191004)%
    \lineheight{1}%
    \setlength\tabcolsep{0pt}%
    \put(0,0){\includegraphics[width=\unitlength,page=1]{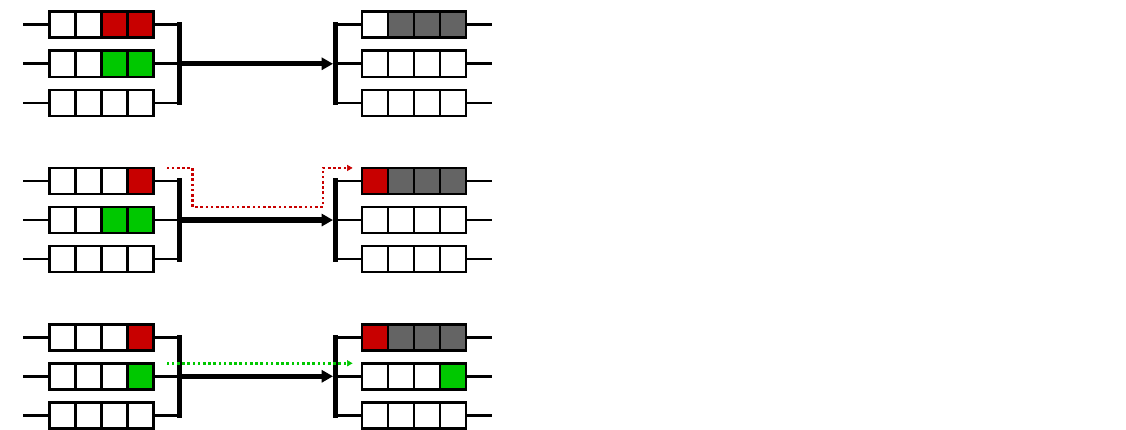}}%
    \put(0.4605278,0.34835443){\color[rgb]{0,0,0}\makebox(0,0)[lt]{\lineheight{0}\smash{\begin{tabular}[t]{l}higher priority flit can move\end{tabular}}}}%
    \put(0.4605278,0.20947559){\color[rgb]{0,0,0}\makebox(0,0)[lt]{\lineheight{0}\smash{\begin{tabular}[t]{l}buffer is now full, packet 1 is blocked\end{tabular}}}}%
    \put(0.4605278,0.04745027){\color[rgb]{0,0,0}\makebox(0,0)[lt]{\lineheight{0}\smash{\begin{tabular}[t]{l}packet 2 can move while 1 is blocked\end{tabular}}}}%
    \put(-0.00240167,0.35992767){\color[rgb]{0,0,0}\makebox(0,0)[lt]{\lineheight{0}\smash{\begin{tabular}[t]{l}${\color{red}1}$\end{tabular}}}}%
    \put(-0.00240167,0.32520796){\color[rgb]{0,0,0}\makebox(0,0)[lt]{\lineheight{0}\smash{\begin{tabular}[t]{l}${\color{green}2}$\end{tabular}}}}%
    \put(0,0){\includegraphics[width=\unitlength,page=2]{NC_model_bypass.pdf}}%
  \end{picture}%
\endgroup%
}
\caption{Bypass mechanism}
\label{fig:bypass}
\end{figure}

The considered wormhole NoC routers are similar to the architecture presented in \cite{kavaldjiev_virtual_channel_router}, illustrated in Figure \ref{fig:router_arch} (top).
They implement a priority-based arbitration of VCs and enable flit-level preemption through VCs. The latter can happen if a flow from a higher priority VC asks for an output that is being used by the flow of interest (\emph{foi}). Hence, when the flit being transmitted finishes its transmission, the higher priority flow is granted the use of the output while the \emph{foi} waits. Moreover, each VC has a specific input buffer and supports many traffic classes, \ie{}, VCs sharing, and many traffic flows may be mapped on the same priority-level, \ie{}, priority sharing. Finally, the implemented VCs enable the bypass mechanism, illustrated in Figure \ref{fig:bypass}. If the \emph{foi} gets blocked at some point (for instance, flow 1 in Figure \ref{fig:bypass}), flows from lower priority VCs sharing upstream outputs with the \emph{foi} (for instance, flow 2 in Figure \ref{fig:bypass}) can bypass it, but they will be preempted again when the downstream blocking of the \emph{foi} disappears.

We consider an arbitrary service policy to serve flows belonging to the same VC within the router, \ie{}, these flows can be from the same traffic class or from different traffic classes mapped on the same VC. This assumption allows us to cover the worst-case behaviors of different service policies, such as FIFO and Round Robin (RR) policies.

Hence, we model such a wormhole NoC router as a set of independent hierarchical multiplexers, where each one represents an output port as shown in Figure \ref{fig:router_arch} (bottom). The first arbitration level is based on a blind (arbitrary) service policy to serve all the flows mapped on the same VC level and coming from different \textit{geographical} inputs. The second level implements a preemptive Fixed Priority (FP) policy to serve the flows mapped on different VCs levels and going out from the same output port. It is worth noticing that the independency of the different output ports is guaranteed in our model, due to the integration of the flows serialization phenomena. The latter induces ignoring the interference between the flows entering a router through the same input and exiting through different outputs, since these flows have necessarily arrived through the same output of the previous router, where we have already taken into account their interference.

Each router-output pair $r$ (that we will refer to as a \emph{node} from now on) has a processing capacity that we model using a rate-latency service curve.
\[ \beta^{r}(t) = R^{r} (t - T^{r})^+ \]
$R^r$ represents the minimal processing rate of the router for this output (which is typically expressed in flits per cycle, fpc) and $T^r$ the maximal experienced delay by any flit crossing the router before being processed (which is commonly called routing delay and takes one or few cycles).

\subsection{Flow Model}
The characteristics of each traffic flow $f \in \mathcal{F}$ are modeled with the following leaky bucket arrival curve, which covers a lot of different traffic arrival events, such as CBR or bursty traffic with or without jitter :
\[ \alpha_f(t) = \sigma_f + \rho_f \cdot t \]
This arrival curve integrates the maximal packet length $L_f$ (payload and header in flits), the period or minimal inter-arrival time $P_f$ (in cycles), the burst (number of packets the flow may release consecutively) $b_f$ and the release jitter $J_f$ (in cycles) in the following way :
\begin{eqnarray*}
  \rho_f   &=& \frac{L_f}{P_f} \\
  \sigma_f &=& b_f \cdot  L_f + J_f \cdot \rho_f
\end{eqnarray*}
If $f$ is CBR flow, we have $b_f=1$.

For each flow $f$, its path $\path_f$ is the list of nodes (router-outputs) crossed by $f$ from source to destination. Moreover, for any $k$ in appropriate range, $\path_f[k]$ denotes the $k+1^{th}$ node of flow $f$ path (starting at index 0). Therefore, for any $r \in \path_f$, the propagated arrival curve of flow $f$ from its initial source until the node $r$, computed based on Th. \ref{TH1}, will be denoted:
\[ \alpha_f^r(t) = \sigma_f^r + \rho_f^r \cdot t \]
The end-to-end service curve granted to flow $f$ on its whole path will be denoted:
\[ \beta_{f}(t) = R_f \left( t - T_f \right)^+ \]

\subsection{Preliminaries}
\label{subsec:preliminaries}
Consider $k$ and $l$ two flows that are directly interfering with one another, $\path_k, \path_l$ their paths, and let $\dv(\path_k, \path_l)$ be the last node they share :
\[ \dv(\path_k, \path_l) = \path_k[ \max \{i, \path_k[i] \in \path_l \}] \]
Suppose the path of $l$ continues after this node. Even if the head flit of $l$ is not stored in a router of $\path_{k} \cap \path_{l}$, the limited buffer size available in each router can lead to storing the tail flit of $l$ in a router of $\path_{k} \cap \path_{l}$ under contention. In that case, $l$ blocks $k$.

Therefore, we need to quantify the way a packet of flow $f$ spreads into the network when it is blocked and stored in buffers. 
Here, we assume node $r$ has a buffer size of $B_r$ to model heterogeneous architectures.
\begin{Definition}
Consider a flow $f$ of maximum packet length  $L_{f}$ flits.
The spread index of $f$ at node $i$, denoted $N_{f}^i$, is defined as follows: 
\[ N_f^{i} = \min_{l \geq 0} \left\{ l, L_f \leq \sum\limits_{j = 0}^{l-1} B^{\path_f[i+j]} \right\} \]
where $B^r$ the buffer size at node $r$ in flits.
\label{def:spread_index}
\end{Definition}
$N_f^i$ is the number of buffers needed to store one packet of flow $f$ from node $\path_f[i]$ onwards on the path of $f$.

Using this notion and the last intuitive example, we call the section of the path of flow $k$ from $\dv(\path_k, \path_l)$ through $N_k^{\dv(\path_k, \path_l)}$ nodes (at most) 
``subpath of $k$ relatively to $l$'':
\begin{Definition}
The subpath of a flow $k$ relatively to a flow $l$ is:
\begin{eqnarray*}
    subpath(\path_k,\path_l) &=& \Big[ \path_k[Last(\path_k,\path_l) + 1 ], \ldots, \\
                             & & \path_l[Last(\path_k,\path_l) + N_k^{Last(\path_k,\path_l) + 1} ] \Big] \\
\end{eqnarray*}
where $Last(\path_k,\path_l) = \max\{ n, \; \path_k[n] \in \path_l\}$ is the index of the last node shared by $k$ and $l$ along $\path_k$, i.e $\path_k[Last(\path_k,\path_l)] = \dv(\path_k,\path_l)$. 
\label{def:subpath1}
\end{Definition}
We can extend this notion and define, in a similar fashion, the subpath of any flow $k$ relatively to a subpath $\subpath_l \subset \path_l$ of any flow $l$ (with $l\neq k$ or $l=k$).
The previous notation still holds:
\begin{Definition}
The subpath of a flow $k$ relatively to any subpath $\subpath_l$ of any flow $l$ is:
\begin{eqnarray*}
    subpath(\path_k, \subpath_l) &=& \Big[ \path_k[Last(\path_k, \subpath_l) + 1 ], \ldots, \\
           & & \path_k[Last(\path_k, \subpath_l) + N_k^{Last(\path_k,\subpath_l) + 1}] \Big] 
\end{eqnarray*}
where $Last(\path_k,\subpath_l) = \max\{ n, \; \path_k[n] \in \subpath_l\}$ is the index along $\path_k$ of the last node shared by $k$ 
and $l$ within $\subpath_l$.
By abuse of notation, we denote $subpath(k,l)$ to refer to $subpath(\path_k,\path_l)$, and similarly $subpath(k, \subpath_l)$ to refer to $subpath(\path_k,\subpath_l)$.
\end{Definition}

\begin{figure}[ht]
\centering
\resizebox{\columnwidth}{!}{
\begingroup%
  \makeatletter%
  \providecommand\color[2][]{%
    \errmessage{(Inkscape) Color is used for the text in Inkscape, but the package 'color.sty' is not loaded}%
    \renewcommand\color[2][]{}%
  }%
  \providecommand\transparent[1]{%
    \errmessage{(Inkscape) Transparency is used (non-zero) for the text in Inkscape, but the package 'transparent.sty' is not loaded}%
    \renewcommand\transparent[1]{}%
  }%
  \providecommand\rotatebox[2]{#2}%
  \newcommand*\fsize{\dimexpr\f@size pt\relax}%
  \newcommand*\lineheight[1]{\fontsize{\fsize}{#1\fsize}\selectfont}%
  \ifx\svgwidth\undefined%
    \setlength{\unitlength}{420.54537459bp}%
    \ifx\svgscale\undefined%
      \relax%
    \else%
      \setlength{\unitlength}{\unitlength * \real{\svgscale}}%
    \fi%
  \else%
    \setlength{\unitlength}{\svgwidth}%
  \fi%
  \global\let\svgwidth\undefined%
  \global\let\svgscale\undefined%
  \makeatother%
  \begin{picture}(1,0.38635563)%
    \lineheight{1}%
    \setlength\tabcolsep{0pt}%
    \put(0,0){\includegraphics[width=\unitlength,page=1]{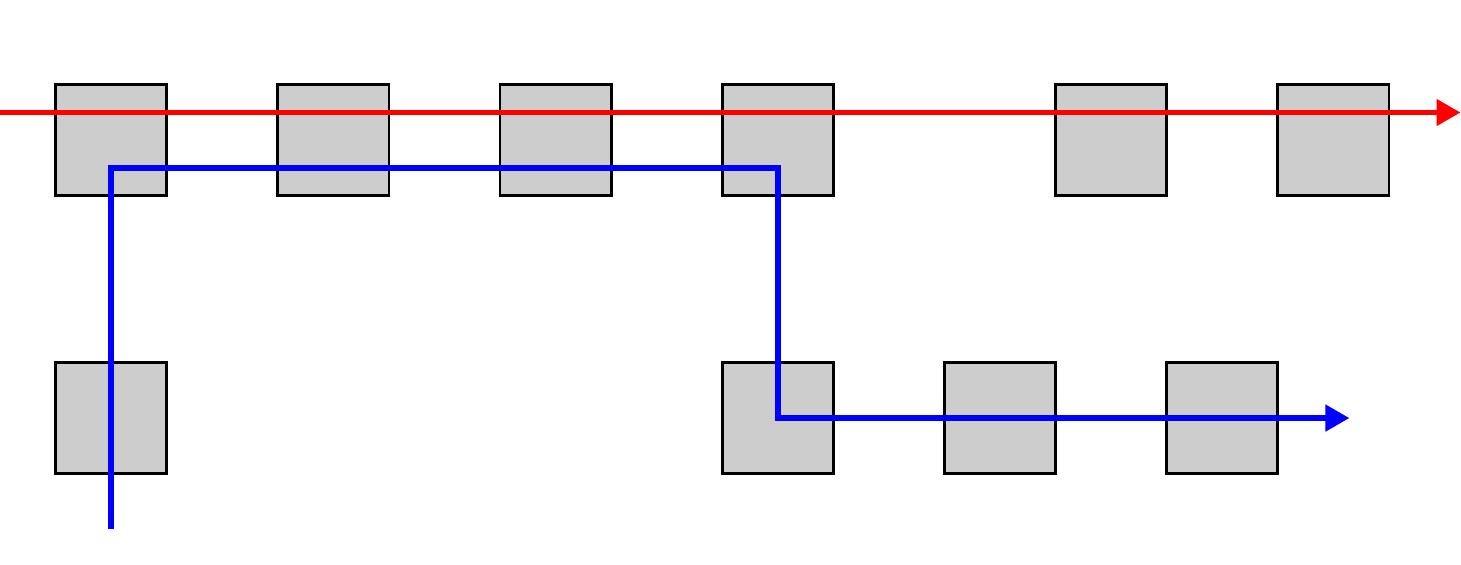}}%
    \put(0,0.27126696){\color[rgb]{0,0,0}\makebox(0,0)[lt]{\lineheight{0}\smash{\begin{tabular}[t]{l}{\color{red} $k$}\end{tabular}}}}%
    \put(0.09511459,0.02396906){\color[rgb]{0,0,0}\makebox(0,0)[lt]{\lineheight{0}\smash{\begin{tabular}[t]{l}{\color{blue} $l$}\end{tabular}}}}%
    \put(0.34241252,0.34735864){\color[rgb]{0,0,0}\makebox(0,0)[lt]{\lineheight{0}\smash{\begin{tabular}[t]{l}$\dv(\path_k, \path_l)$\end{tabular}}}}%
    \put(0,0){\includegraphics[width=\unitlength,page=2]{NC_model_subpath.pdf}}%
    \put(0.49459586,0.00494614){\color[rgb]{0,0,0}\makebox(0,0)[lt]{\lineheight{0}\smash{\begin{tabular}[t]{l}$subpath(l,k)$\end{tabular}}}}%
    \put(0.03804584,0.34735865){\color[rgb]{0,0,0}\makebox(0,0)[lt]{\lineheight{0}\smash{\begin{tabular}[t]{l}$\conv(\path_k,\path_l)$\end{tabular}}}}%
    \put(0.49459586,0.34735865){\color[rgb]{0,0,0}\makebox(0,0)[lt]{\lineheight{0}\smash{\begin{tabular}[t]{l}$\path_l[Last(\path_l,\path_k) + 1]$\end{tabular}}}}%
  \end{picture}%
\endgroup%
}
\caption{Subpath illustration for the \emph{foi} $k$}
\label{fig:subpath}
\end{figure}

It is worth noticing that if $\path_l$ ends before reaching the $N_l^{Last(\path_k,\path_l) + 1}$-th node after $\dv(\path_k,\path_l)$, then we ignore the out-of-range indexes. The notion of subpath is illustrated in Figure \ref{fig:subpath} for the \emph{foi} $k$ and a spread index for the interfering flow $l$ right after node $\dv(\path_k, \path_l)$ equal to 3, \ie{}, $N_l^{Last(\path_k, \path_l) + 1} =3$.

We will also need the following definitions.
\begin{Definition}
Let $f$ be the \emph{foi}. The set of flows that share resources with $f$ on their paths is called the Direct Blocking set of $f$ and denoted $DB_f$. Moreover, the subset of flows in $DB_f$ sharing resources with $f$ along $path$ is denoted $DB_f^{|path}$.  
\end{Definition}
\begin{Definition}
Let $f$ be the \emph{foi}. 
$hp(f)$ is the set of flows mapped to a VC of strict higher priority than $f$.
$sp(f)$ is the set of flows mapped to the same VC as $f$, $f$ excluded.
$lp(f)$ is the set of flows mapped to a VC of strict lower priority than $f$.
Morevoer, we define $slp(f) = sp(f) \cup lp(f)$ (resp. $shp(f) = sp(f) \cup hp(f)$), that is all flows with a priority lower or equal (resp. higher or equal) than $f$, $f$ excluded.
\end{Definition}

\begin{Definition}\label{def:ibset}
The indirect blocking set of a flow $f$ is the set of flows that do not physically share any resource with the \emph{foi}, but cause a delay to the \emph{foi} because they impact (directly or indirectly) at least one flow of $DB_f$. 
It is denoted $IB_f$ and contains pairs of the form \{flow id, subpath\} to specify, for each flow, where a packet of that flow can cause blocking that may propagate to the \emph{foi} through backpressure.
\end{Definition}

It is worth noticing that Definition \ref{def:ibset} is slightly different from the one used in the Scheduling Theory approaches \cite{burns_priority_share} \cite{liu_tighter_time_analysis}, where there is a distinction between the indirect blocking, due to same-priority flows, and indirect interference, due to higher priority flows. In our approach, we only consider flows belonging to the same VC as the \emph{foi} to compute the indirect blocking set, since the impact of higher priority flows is already integrated in our model as follows:
\begin{itemize}
    \item if a higher priority flow blocking our \emph{foi} gets blocked, the \emph{foi} can bypass it. In this case, we take into account the extra processing delay needed to allocate the shared resource to the \emph{foi}. On the other hand, the buffer backpressure will only propagate among flows from the same class, as illustrated in Figure \ref{fig:bypass};
    \item the influence of higher priority flows on the same priority flows than the \emph{foi}, which are inducing the indirect blocking, is modeled through the granted end-to-end service curve of each one of these flows at the rate and latency levels, as explained in Section \ref{subsec:direct_blocking_latency}
\end{itemize}

%

\section{Graph-based Approach for Buffer-Aware Timing Analysis}
\label{sec:interference_graph_approach}
%

%

Hereafter, we propose a graph-based approach to compute the IB set of flows to cover CPQ scenarios in Section \ref{subsec:ibset_graph}. 
Then, we detail the new method to compute $T_{IB}$ in Section \ref{subsec:ib_latency_graph}.
We illustrate each step on an example.

We first present an overview of G-BATA with the needed steps to compute the end-to-end delay bound (\ref{subsec:main_steps}). 
In the following sections (\ref{subsec:direct_blocking_latency} to \ref{subsec:ib_latency_graph}), we detail each step and illustrate them with an example. 

%

\subsection{Overview}
\label{subsec:main_steps}

To get a bound on the end-to-end latency for a flow $f$, we first need to compute its end-to-end service curve.
The end-to-end service curve of $f$ is denoted:
\[ \beta_{f}(t) = R_f \left( t - T_f \right)^+ \]
where $T_f$ is the sum of:
\begin{itemize}
    \item $T_{\path_f}$, the ``base latency'', that any flit of $f$ experiences along its path due only to the technological latencies of the crossed routers; 
    \item $T_{DB}$, the maximum direct blocking latency, due flows in $DB_f$;
    \item $T_{IB}$, the maximum indirect blocking latency, due to flows in $\IB_f$. 
\end{itemize}



To compute the bound on the end-to-end latency for the \emph{foi} $f$, we proceed according to Algorithm \ref{alg:end_to_end_service_computation} and following these main steps: 
\begin{enumerate}
    \item We compute $T_{\path_f}$ (Line \ref{lin:T_P_f}), and the direct blocking latency $T_{DB}$ (Lines \ref{lin:T_DB_start} to \ref{lin:T_DB_end});
    \item We compute the indirect blocking set $\IB_f$ (Line \ref{lin:IB_f});
    \item We compute $T_{IB}$ (Lines \ref{lin:T_IB_start} to \ref{lin:T_IB_end}). 
    \item From there on, knowing the initial arrival curve of $f$, $\alpha_f$, and its end-to-end service curve $\beta_f$ (Line \ref{lin:beta_f}), we compute the end-to-end delay bound using Theorem \ref{TH1} as follows:
\begin{equation}
\label{eed}
D_f^{\path_f} = \frac{\sigma^{\path_f[0]}}{R_f} + T_{hp} + T_{sp} + T_{lp} + T_{IB} + T_{\path_f} 
\end{equation}
where $\sigma^{\path_f[0]}$ is the burst of the initial arrival curve of $f$ (the arrival curve of $f$ at node $\path_f[0]$, \ie{} the first node of its path).
\end{enumerate}

The main steps that are impacted by CPQ scenarios are steps 2 and 3. 
The remaining steps are the same as with BATA approach, introduced in \cite{giroudot_buffer_aware}. 
Therefore, we recall herein only the main idea of step 1 for self-containment purpose and more details can be found in \cite{giroudot_buffer_aware}, and we rather focus on the details of steps 2 and 3 illustrating the introduced graph-based approach to cope with the CPQ assumption.

\begin{algorithm}
\caption{Computing the end-to-end service curve for a flow $f$}
\label{alg:end_to_end_service_computation}
endToEndServiceCurve($f, \path_f$)
\begin{algorithmic}[1]
    \STATE Compute $R_f$
    \STATE Compute $T_{\path_f}$ \label{lin:T_P_f}\\

    \COMMENT{Compute $T_{DB}$:}
    \STATE $T_{DB} \leftarrow 0$ 
    \FOR{$k \in DB_f$}\label{lin:T_DB_start}
        \STATE $r_0 \leftarrow \conv(k,f)$ \COMMENT{Get convergence point of $f$ and $k$}
        \STATE $\beta_k \leftarrow $ endToEndServiceCurve($k, [\path_k[0], \cdots, r_0]$)\label{lin:recursive_db}
        \STATE $\alpha_k^0 \leftarrow $ initial arrival curve of $k$
        \STATE $\alpha_k \leftarrow \alpha_k^0 \oslash \beta_k$
        \STATE $T_{DB} \leftarrow $ directBlocking($\alpha_k^{r_0}$)
    \ENDFOR\label{lin:T_DB_end}

    \COMMENT{Compute $T_{IB}$:}
    \STATE Compute $\IB_f$  \label{lin:IB_f}
    \STATE $T_{IB} \leftarrow 0$\label{lin:T_IB_start}
    \FOR{$\{k,S\} \in IB_f$}
        \STATE $\widetilde{\beta}_k \leftarrow$ VC-service curve of $k$ on $S$\\
        \STATE $\alpha_k \leftarrow$ initial arrival curve of $k$ \label{lin:subpath_arrival_curve}\\
        \COMMENT{Now add the latency over the subpath to $T_{IB}$ :}
        \STATE $T_{IB} \leftarrow T_{IB} + h(\alpha_k, \widetilde{\beta}_k) $
    \ENDFOR\label{lin:T_IB_end}

    \RETURN $\beta = R_f (t - (T_{\path_f} + T_{DB} + T_{IB}))^+$\label{lin:beta_f}
\end{algorithmic}
\end{algorithm}

\subsection{Step 1: Direct Blocking Latency Computation}
\label{subsec:direct_blocking_latency}

The direct blocking latency takes into account the impact of flows sharing resources with the \emph{foi}. 
We used PMOO \cite{schmitt_pmoo} to account for flow serialization phenomena when computing the maximum direct blocking latency.
As introduced in \cite{giroudot_buffer_aware}, it is defined in the following Theorem.
\begin{Theorem}
\label{Th:directblocking}(Maximum Direct Blocking Latency)\\
The maximum direct blocking latency for a \emph{foi} $f$ along its path $\path_f$,  in a NoC under flit-level preemptive FP multiplexing with strict service curve nodes of the rate-latency type $\beta_{R,T}$ and leaky bucket constrained arrival curves $\alpha_{\sigma, \rho}$ is equal to:
 \[ T_{hp} + T_{sp} + T_{lp}\] 

with:
\begin{subequations}
\label{directBlocking}
\begin{align}
&  T_{hp} =  \sum\limits_{i \in DB_f \cap hp(f)} \frac{\sigma_i^{\conv(i,f)} + \rho_i \cdot \sum\limits_{r \in \path_f \cap \path_i} \left( T^r + \frac{ L_{slp(f)}^{r} }{R^{r}} \right)}{R_f}  \\
&  T_{sp} =  \sum\limits_{i \in DB_f \cap sp(f)} \frac{\sigma_i^{\conv(i,f)} + \rho_i \cdot \sum\limits_{r \in \path_f \cap \path_i} \left( T^r + \frac{ L_{slp(f)}^{r} }{R^{r}} \right)}{R_f}  \\
& T_{lp} = \sum\limits_{r \in \path_f} \frac{L_{slp(f)}^{r} }{R^{r}} 
\end{align}
\end{subequations}

where:
\begin{subequations}
\label{directBlockingvariables}
\begin{align}
&  L_{slp(f)}^{r} = \max\left( \max\limits_{j \in sp(f)} \left( L_{j} \cdot \mathbf{1}_{\{sp(f) \supset r\}} \right) \, , S_{flit} \cdot \mathbf{1}_{\{lp(f) \supset r \}} \right) \nonumber \\
&  R_f =  \min\limits_{r \in \path_f} \left\{ R^r - \sum\limits_{j \ni r, j \in shp(f)} \rho_j \right\} \nonumber
\end{align}
\end{subequations}
\end{Theorem}
The proof can be found in \cite{giroudot_buffer_aware}. 

\begin{application}
    We now detail the computations on the example of Figure \ref{fig:problematic_example_cpq}, for the \emph{foi} 1. 
    We assume all routers have a service curve $\beta = R (t - T)^+$ and flow $i$ has a packet length $L_i = L$ and the initial arrival curve $\alpha_i = \sigma + \rho t$. We also consider all flows have a burst $b = 2$ and no jitter, therefore $\sigma = 2L$. 
    All flows are mapped to the same VC, thus $T_{hp} = T_{lp} = 0$. 
    We then have :
    \begin{eqnarray*}
	    T_{\path_1} &=& 4 T\\
	    T_{sp}      &=& \frac{\sigma_2^{R3} + \rho (\cdot T + \frac{L_i}{R})}{R - \rho} \\
	    &=& \frac{\sigma + \rho \cdot (T + \frac{L}{R})}{R - \rho} \\
    \end{eqnarray*}

    Hence :
    \begin{eqnarray*}
    T_{DB} &=& 4 T + \frac{2 L}{R - \rho} + \rho\frac{ T + \frac{L}{R} }{R - \rho}\\
           &=& 10,526315789 \textrm{ cycles}
    \end{eqnarray*}
    with $R = 1$ flit/cycle, $T = 1$ cycle, $\rho = 0.05$ flits/cycle, $\sigma = 3$ flits and $L = 3$ flits.
\end{application}

It is important to notice that when we compute the direct blocking latency $T_{DB}$ of the \emph{foi}, we need to know the burst of interfering flows at their convergence point with the \emph{foi}.
Thus we need to compute, for each interfering flow, its service curve from its source to the aforementioned convergence point.

The end-to-end service curve computation is thus a recursive process (cf. Algorithm \ref{alg:end_to_end_service_computation}).
The recursion terminates because each call to endToEndServiceCurve() is done upstream the current convergence point.

\subsection{Step 2: Indirect Blocking Set Computation}
\label{subsec:ibset_graph}

To handle CPQ assumption,
we start from two modifications.
First, we allow to compute the subpath of any flow $f$ relatively to a subpath $\subpath_f \subset \path_f$ of $f$ to model several packets of the same flow queuing in the network.
Second, we use a graph structure to maintain the dependency information between the subpaths.
By doing so, we are able to know how each subpath was computed, and we also can explore all possible interference patterns more easily.

Each vertex corresponds to a subpath of a flow and holds the following information :
\begin{itemize}
    \item \textbf{fkey} : the flow identifier
    \item \textbf{path} : the subpath
    \item \textbf{dependencies} : the list of all edges $(v, u)$ where $v$ is the current vertex and $u$ is such that $v.path$ is the subpath of flow $v.fkey$ relatively to subpath $u.path$.
    \item \textbf{dependents} : the list of all edges $(w, v)$ where $v$ is the current vertex and $w$ is such that $w.path$ is the subpath of flow $w.fkey$ relatively to subpath $v.path$. 
\end{itemize}

The two functions to construct the graph are detailed in Algorithm \ref{alg:ibgraph} and \ref{alg:next_vertices}.
The main steps are as follows:
\begin{enumerate}
  \item We create a graph with one vertex corresponding to the \emph{foi} (Line \ref{lin:init_graph});
  \item We compute all subpaths relatively to the \emph{foi} and create a vertex depending on the \emph{foi}'s vertex for each non-empty subpath (Lines \ref{lin:init_list_next_vertices} and \ref{lin:update_list_next_vertices});
  \item We add these vertices to the graph, making sure there are no dupplicates and merging the dependencies of the new vertex with the existing one if needed (Line \ref{lin:add_vertices});
  \item We iterate these steps on each new vertex, in a breath-first manner, until no new vertex is created (loop on Line \ref{lin:loop}). 
\end{enumerate}

Once the graph is created, the indirect blocking set of $f$ consists of the pairs $(k, subk)$ from all vertices such that $k \notin DB_f \cup \{f\}$. 
In other words, these vertices correspond to flows that do not directly interfere with the \emph{foi} $f$.

\begin{algorithm}
\caption{Computing the indirect blocking graph for flow $f$}
\label{alg:ibgraph}
constructIBGraph()
\begin{algorithmic}[1]
    \REQUIRE{$f$, the flow of interest, $\path_f$ the associated path, $\mathcal{F}$ the set of flows}
    \ENSURE{$\ibg_f$, a graph of all subpaths involved in indirect blocking patterns impacting $f$}

    \STATE $v_0 \leftarrow $ vertex($f, \path_f, [], []$) \label{lin:init_graph}
    \STATE $\mathcal{L}_0 \leftarrow$ getNextVertices($[v_0], \mathcal{F}$) \COMMENT{Initialize a list}\label{lin:init_list_next_vertices}

    \WHILE{$\mathcal{L}_0 \neq []$}\label{lin:loop}
        \FOR{$v \in \mathcal{L}_0$}
            \STATE addVertex($\ibg_f, v$) \label{lin:add_vertices}
        \ENDFOR
        \STATE $\mathcal{L}_0 \leftarrow$ getNextVertices($\mathcal{L}_0, \mathcal{F}$)\label{lin:update_list_next_vertices}
    \ENDWHILE

    \RETURN{$\ibg_f$}
\end{algorithmic}
\end{algorithm}

\begin{algorithm}
\caption{Computing vertices and adding them to the graph}
\label{alg:next_vertices}
getNextVertices($\mathcal{L}_{in}$, $\mathcal{F}$)
\begin{algorithmic}[1]
    \REQUIRE{$\mathcal{L}_{in}$ a list of vertices, $\mathcal{F}$ the flowset}
    \ENSURE{$\mathcal{L}_{out}$, a list of the vertices depending on the vertices of $\mathcal{L}_{in}$}
    \FOR{$v \in \mathcal{L}_{in}$}
        \FOR{$k \in \mathcal{F}$}
            \STATE $\subpath \leftarrow subpath(k, v.path)$
            \IF{$\subpath \neq \varnothing$}
                \STATE $w \leftarrow$ vertex($k, \subpath, [v], []$)
                \STATE append $w$ to $\mathcal{L}_{out}$
            \ENDIF
        \ENDFOR
    \ENDFOR
    \RETURN $\mathcal{L}_{out}$
\end{algorithmic}
%
addVertex($\ibg, v$)
\begin{algorithmic}[1]
    \IF{$\exists w \in \ibg$ such that $w.path = v.path$ and $w.fkey = v.fkey$}
        \STATE merge $v$ with $w$
    \ELSE
        \STATE add $v$ to $\ibg$
    \ENDIF
\end{algorithmic}
\end{algorithm}

The computational complexity of Algorithm \ref{alg:ibgraph}, when considering a flow set $\mathcal{F}$ on the NoC, is denoted as $\mathcal{C}(|\mathcal{F}|)$ and is defined in the following property.
\begin{Property}
\label{prop:complexity_graph}
Consider a flow set $\mathcal{F}$, the computational complexity of Algorithm \ref{alg:ibgraph} is as follows:
\begin{equation}
\label{eq:complexity_graph}
 \mathcal{C}(|\mathcal{F}|) = \mathcal{O}\left( \max\limits_{f \in \mathcal{F}} |\path_f| \cdot \sum\limits_{f \in \mathcal{F}} |\path_f| \right)
\end{equation}
and can be roughly bounded as follows :
\begin{equation}
\label{eq:rough_complexity_bound}
\mathcal{C}(|\mathcal{F}|) = \mathcal{O} \left( (\max\limits_{f \in \mathcal{F}} |\path_f|)^2 \cdot |\mathcal{F}| \right)
\end{equation}

\end{Property}

\begin{IEEEproof}
We first notice that vertices of the graphs are defined only by their flow index and subpath.
For a flow $f$, there are $|\path_f|$ possible subpaths (each of them starting at a different node of the path of $f$). Therefore, there are at most $\sum_{f \in \mathcal{F}} |\path_f|$ distinct subpaths for the flowset $\mathcal{F}$.

We can thus bound the number of vertices of the computed graph. For each of these vertices, the algorithm computes all possible subpaths relatively to the current vertex' subpath (in getNextVertices() main loop).

Assume this subpath is $S$ and that we have a preprocessed dictionary listing, for every node, the indexes of flows using this node,\footnote{We do run such a preprocessing on the configuration.}.
Although we wrote the secondary loop of getNextVertices() as a loop over all flows in $\mathcal{F}$ for clarity reasons, all we have to do to get all possible subpaths relatively to $S$ is run through the nodes of $S$ and check for intersection with another flow's path.
Comparing the indexes of the current node with those of the previous node, we can find divergence nodes of contending flows relatively to $S$.
We assume that, knowing the divergence point of a contending flow relatively to $S$, it takes a constant time to find its subpath (we only need to compute the spread index).

Thus, the complexity of finding all subpaths relatively to any subpath is $\mathcal{O}(\max_{f \in \mathcal{F}} |\path_f|)$, hence the final result.
The last bound is found bounding each path length of the sum by the maximal path length in the whole flow set.
\end{IEEEproof}

The reason we can account for more than one packet of the same flow stalling in the network is because we allow to compute the subpath of a flow relatively to a subpath of that very same flow.
%
%

\begin{application}
    We now apply the algorithm to the configuration of Figure \ref{fig:problematic_example_cpq}: 
    \begin{enumerate}
	    \item starting from flow 1, we create vertex $v_1$ with index 1 and path $\path_1$ and we call getNextVertices() on $[v_1]$. We get $v_2 = {\rm vertex}(2, \subpath_b)$. Since $v_2$ was computed from $v_1$, we add $v_2$ in $v_1.dependents$ and $v_1$ in $v_2.dependencies$. 
	    \item we call getNextVertices() on $[v_2]$. We get $v_2' = {\rm vertex}(2, \subpath_b)$, add $v_2'$ in $v_2.dependents$ and $v_2$ in $v_2'.dependencies$; 
        \item we call getNextVertices() on $[v_2]$. We get $v_3 = {\rm vertex}(3, \subpath_c)$;
        \item we call getNextVertices() on $[v_3]$. We get $v_3' = {\rm vertex}(3, \subpath_d)$.
	\item we call getNextVertices() on $[v_3']$. It returns the empty list [] and the algorithm terminates.
    \end{enumerate}
    The subpaths corresponding to the computed vertices are represented on Figure \ref{fig:subpath_computation_graph}.
    The final graph is the following:\\
\begin{tikzpicture}
    \tikzstyle{every node} = [circle, fill=gray!30]
    \node (1)  at (0, 2) {$1:\path_1$};
    \node (2a) at (2, 1) {$2:\subpath_a$};
    \node (2b) at (4, 1) {$2:\subpath_b$};
    \node (3a) at (6, 1) {$3:\subpath_c$};
    \node (3b) at (8, 1) {$3:\subpath_d$};
    \foreach \from/\to in {1/2a, 2a/2b, 2b/3a, 3a/3b}
    \draw [<-] (\from) -- (\to);
\end{tikzpicture}\\
    and the associated IB set :
    \[IB_1 = \left\{ \{3,\subpath_c\}, \{3, \subpath_d\} \right\}\]
    \begin{figure}[ht]
        \centering
        \resizebox{0.7\columnwidth}{!}{\input{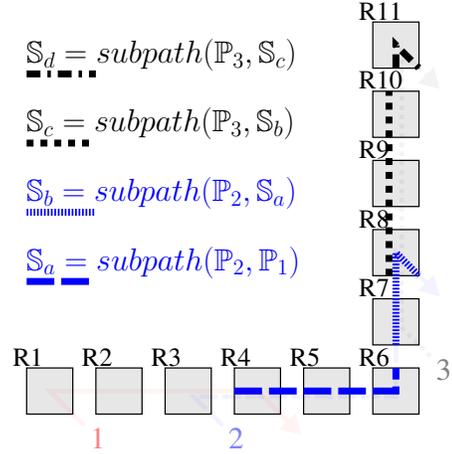}}
        \caption{Subpaths computation with G-BATA approach}
        \label{fig:subpath_computation_graph}
    \end{figure}
\end{application}

\subsection{Step 3: Indirect Blocking Latency Computation}
\label{subsec:ib_latency_graph}

When using G-BATA approach, we take into account the possible queueing of several packets of each flow through the consideration of multiple consecutive subpaths for one flow.
Therefore, when computing $T_{IB}$, the main difference compared to the BATA approach is that, for each \{flow index, subpath\} pair of the derived IB set, we do not need to compute the arrival curve at the beginning of the subpath and instead use the initial arrival curve of one packet of the corresponding flow.
Having several consecutive subpaths for the same indirectly interfering flow allows to take into account a burst of more than one packet.


Computation of the indirect blocking latency $T_{IB}$ is done using the following Theorem :
\begin{Theorem}
\label{thm:t_ib_graph}(Maximum Indirect Blocking Latency)\\
The maximum indirect blocking latency for a \emph{foi} $f$ along its path $\path_f$, in a NoC under flit-level preemptive FP multiplexing with strict service curve nodes of the rate-latency type $\beta_{R,T}$ and leaky bucket constrained arrival curves $\alpha_{\sigma, \rho}$, is as follows:
\begin{equation}
\label{eqn:t_ib_graph}
T_{IB} = \sum_{\{k,subP\} \in IB_f} \frac{ L_k + J_k\rho_k }{\widetilde{R}_k^{subP} } + \widetilde{T}_k^{subP}
\end{equation}
where:
\begin{subequations}
\label{eqn:vc_scurve_graph}
\begin{align}
& \widetilde{R}_k^{subP} =  \min\limits_{r \in subP} \left\{ R^r - \sum\limits_{j \ni r, j \in hp(f)} \rho_j \right\} \label{eqn:vc_scurve_R_graph}  \\
& \widetilde{T}_k^{subP} =  \sum\limits_{r \in subP} \left( T^r
                                        +  \frac{S_{flit} \mathbf{1}_{ \{lp(k) \supset r\}} }
                                                 {R^{r}} \right)  +  \nonumber \\
&    \sum\limits_{i \in DB_k^{|subP} \cap hp(k)} \frac{\sigma_i^{\conv(i,k)} +
           \rho_i \sum\limits_{r \in subP \cap \path_i}
           \left( T^r + \frac{S_{flit} \mathbf{1}_{\{lp(k) \supset r\}} }{R^{r}} \right) }{\widetilde{R}_k^{subP}} \label{eqn:vc_scurve_T_graph}
\end{align}
\end{subequations}
\end{Theorem}

\begin{IEEEproof}
For any pair $\{j, subP_j\} \in IB_f$, a packet of flow $j$ will impact the \emph{foi} $f$ during the maximum time it occupies the associated subpath $subP_j$, $\Delta t_j^{max}$. Hence, a safe upper bound on the indirect blocking latency is as follows:
\[ T_{IB} \leq \sum\limits_{\{j, subP_j\} \in IB_f}  \Delta t_j^{max} \]

On the other hand, for any pair $\{j, subP_j\} \in IB_f$, $\Delta t_j^{max}$ is upper bounded by the end-to-end delay bound of one packet of flow $j$ along its associated subpath $subP_j$, $D_j^{subP_j}$, which infers the following:
\begin{equation}
\label{T-IB}
T_{IB} \leq \sum\limits_{\{j, subP_j\} \in IB_f}  D_j^{subP_j}
\end{equation}

Based on Theorem \ref{TH1}, the delay bound of flow $j$, $D_j^{subP_j}$, is computed as the maximum horizontal distance between:
\begin{itemize}
\item the maximum arrival curve for a single packet of flow $j$ at the input of the subpath $subP_j$, $\alpha_j^{subP_j[0]}$. We consider one packet per subpath. This is due to the fact that each subpath holds one packet (from the definition of the spread index). The multiple number of packets is taken into account through the multiple consecutive subpaths of the the same flow. Thus, the considered arrival curve is the initial arrival curve of flow $j$ with $b_j$ equal to one, that is with a burst equal to $L_j + J_j\rho_j$;
\item the granted service curve to flow $j$ by its VC along $subP_j$, $\widetilde{\beta}_j^{subP_j}$, called VC-service curve, when ignoring the same-priority flows (which are already included in $\IB_f$). The latter condition is due to the pipelined behavior of the network, where the same-priority flows sharing $subP_j$ are served one after another if they need shared resources. Hence, the impact of the same-priority flows than flow $j$ is already integrated within the sum expressed in Eq. (\ref{T-IB}).
\end{itemize}

To compute the granted service curve $\widetilde{\beta}_j^{subP_j}$ for each flow $j \in IB_f$ along $subP_j$, 
we apply the existing Theorem \ref{Th:pmoo_fp}, when:
\begin{itemize}
  \item ignoring the same-priority flows in $sp(j)$, thus all $shp(j)$ will become $hp(j)$ and $slp(j)$ will become $lp(j)$ in Eqs. (\ref{pmoo_fp} a) and (\ref{pmoo_fp} b);
  \item considering the flit-level preemption, thus the impact of lower-priority flows in Eq.  (\ref{pmoo_fp} a) is bounded by the maximum transmission time of $S_{flit} \cdot \mathbf{1}_{\{lp(k) \supset r \}}$ within each crossed node $r \in subP_j$;
  \item considering only the direct blocking flows of $j$ 
	  along $subP_j$, thus considering $DB_j^{|subP_j} \cap hp(j)$ in Eq. (\ref{pmoo_fp} b).
\end{itemize}

Hence, we obtain $\widetilde{R}_j^{subP_j}$ and $\widetilde{T}_j^{subP_j}$ described in Eqs. (\ref{eqn:vc_scurve_graph}a) and (\ref{eqn:vc_scurve_graph}b), respectively. Consequently, the maximum indirect blocking latency in Eq. (\ref{T-IB}) can be re-written as follows:
\begin{equation}
\label{T-IB-bis}
T_{IB} \leq \sum\limits_{\{j,subP_j\} \in IB_f} \frac{L_j + J_j\rho_j }{\widetilde{R}_j^{subP_j} } + \widetilde{T}_j^{subP_j}
\end{equation}
\end{IEEEproof}

It is worth noticing that compared to the BATA approach, we do not need to propagate the arrival curves of flows in $\IB_f$ at the beginning of the subpaths when computing $T_{IB}$.
Consequently, our new approach G-BATA does not need to compute service curves upstream the subpaths, which decreases the number of recursive calls to endToEndServiceCurve() in Algorithm \ref{alg:end_to_end_service_computation}.
We will evaluate the associated complexity gain in our computational analysis in Section \ref{subsec:computational_gbata}.

\section{Performance Evaluation}
\label{sec:performance_eval}
In this section, we first analyse the computational effort of G-BATA and particularly on heavy configurations, with reference to BATA. Afterwards, we conduct a sensitivity analysis of the proposed approach when varying the system parameters and analyse their effect on the end-to-end delay bound. Finally, we assess the tightness of the derived bounds, using the insights we got from the sensitivity analysis. 

\subsection{Computational Analysis}
\label{subsec:computational_gbata}

In this section, we study the computational aspect of G-BATA. 
We will first run G-BATA on configurations with 4, 8, 16, 32, 48, 64, 80, 96 and 128 flows on a $8\times8$ NoC and compare it with BATA. 
We randomly generated 20 such configurations for each number of flows $N$.
To do so, we randomly pick $2N$ ($x$-coordinate, $y$-coordinate)-couples, where each coordinate is uniformly chosen in the specified range (here, from 0 to 7).
We use $N$ of these couples for source cores and the other $N$ for destination cores.
There are 20 configurations for each flow number, and we set a time limit of two hours for the analysis. 

\begin{figure*}[ht]
\centering
\includegraphics[scale=0.55]{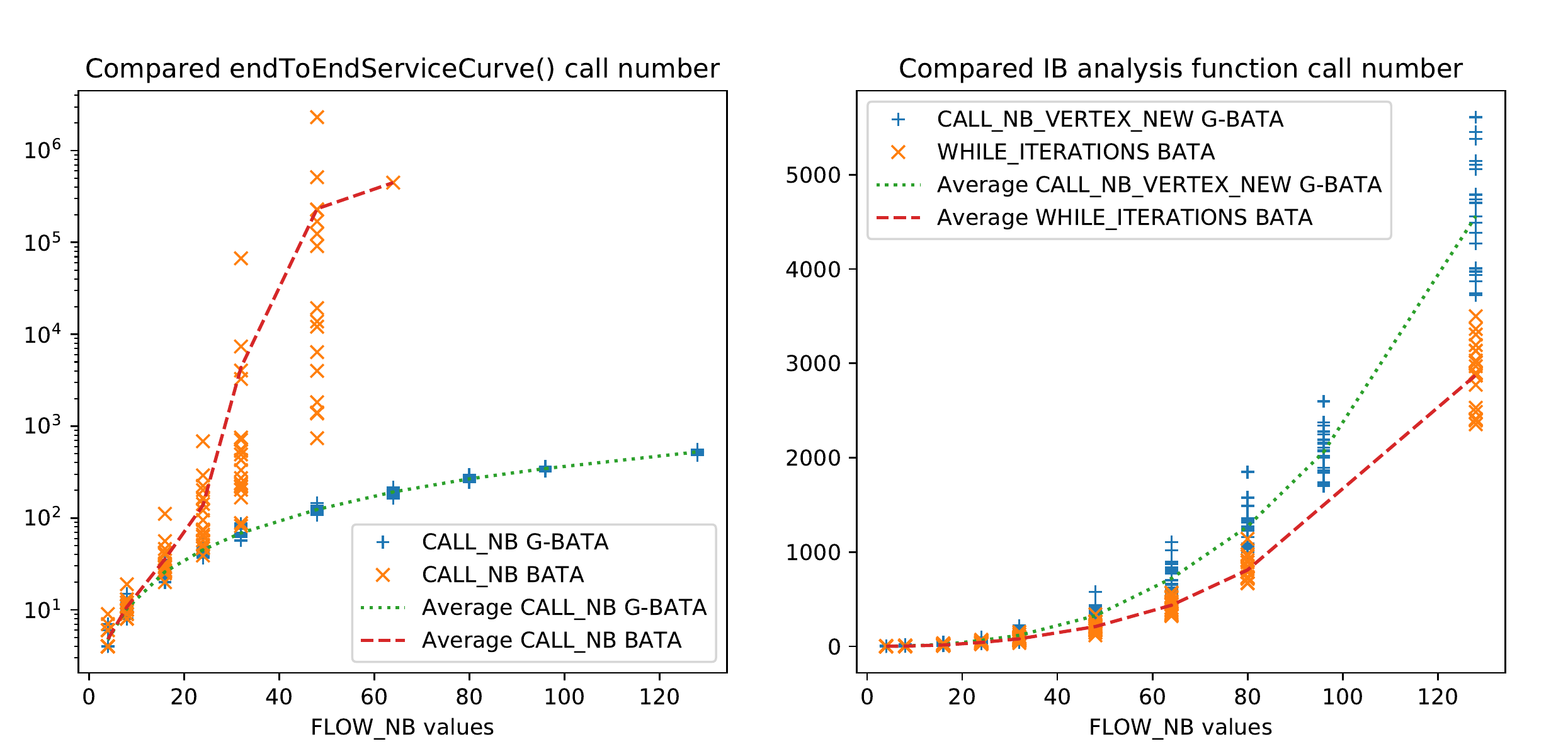}
\caption{Comparative study of the algorithmic complexity}
\label{fig:compared_call_nb}
\end{figure*}

For each configuration, we will focus on the following \emph{complexity metrics}, that give an idea of the cost of analyzing a configuration:
\begin{itemize}
    \item $\Delta t$, the total analysis runtime;
    \item $\Delta t_{IB}$, the duration of the IB analysis (for BATA, determining IB set; for G-BATA, constructing the interference graph);
    \item $\Delta t_{e2e}$, the duration of all end-to-end delay bounds computation;
    \item $N_{e2e}$, the number of calls to the function \verb+endToEndServiceCurve()+;
    \item $N_{iter}$, the number of calls to a representative IB analysis function:
        \begin{itemize}
            \item for BATA, the number of iterations needed to compute all subpaths in the IB set (denoted \emph{while} iterations on Figure \ref{fig:compared_call_nb});
      \item for G-BATA, the number of calls to the function \verb+getNextVertices()+;
        \end{itemize}
\end{itemize}

\begin{figure*}[ht]
\centering
\includegraphics[scale=0.55]{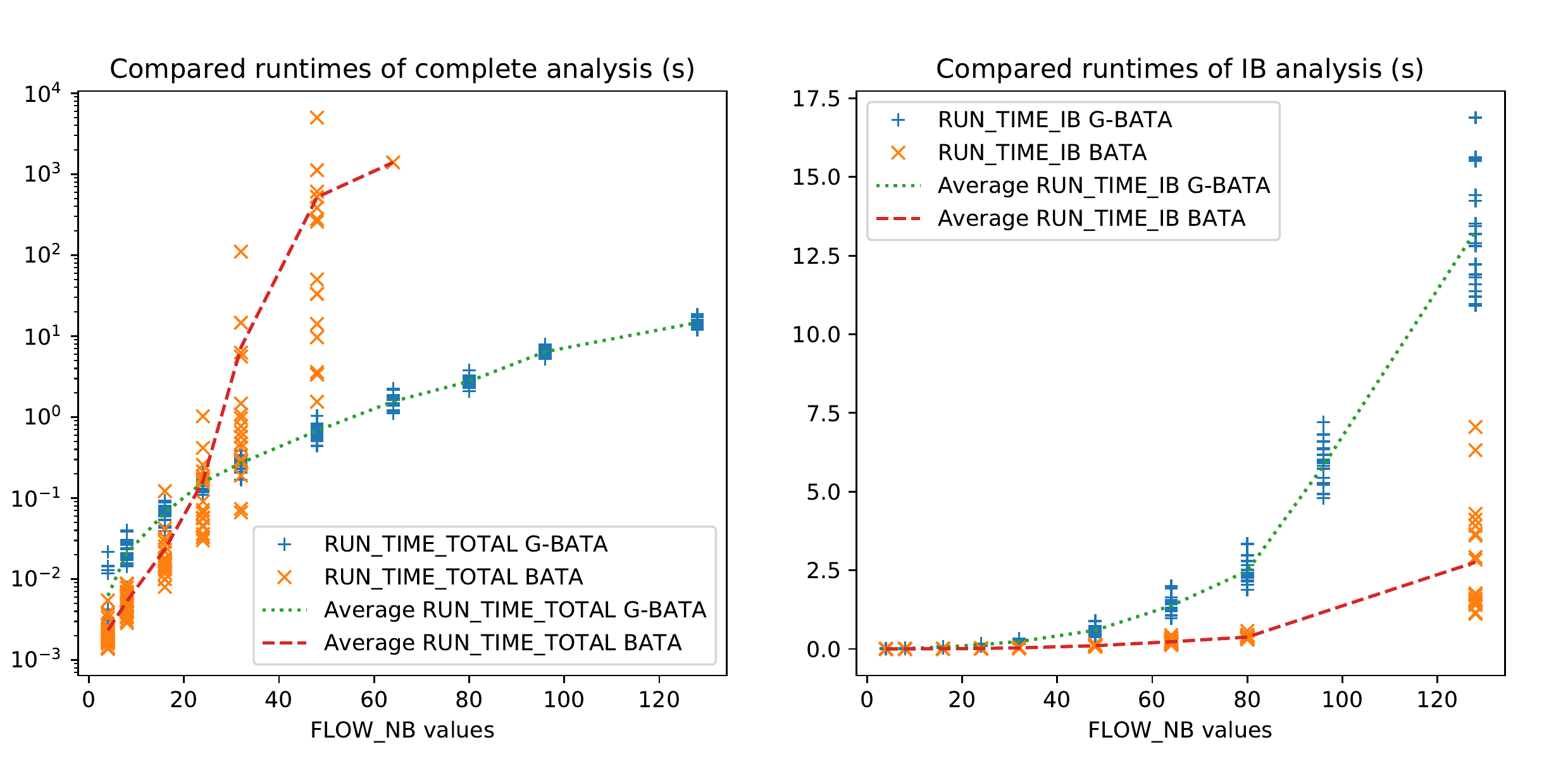}
\caption[Compared runtimes of BATA and G-BATA approaches]{Compared runtimes of both approaches: total runtimes (left) and IB analysis runtimes (right)}
\label{fig:compared_runtimes}
\end{figure*}

We begin the comparative study by plotting the total analysis runtime $\Delta t$ as well as the duration of IB analysis $\Delta t_{\IB}$ as a function of the number of flows in the configuration (Figure \ref{fig:compared_runtimes}).
The first thing we can notice, on the left graph, is that BATA takes more time than G-BATA, especially for flow sets of more than 32 flows.
For instance, the total analysis of 48-flow configurations is on average 766 times faster with G-BATA than with BATA. 
There were no timeouts for G-BATA, whereas BATA timed out for most configurations with 64 flows or more. 

However, we expect the IB analysis part of BATA approach to be computationally less expensive than G-BATA. Since the IB analysis is independent from the end-to-end service curve and delay bound computation, we were able to do it with no time-outs. 
We have plotted the runtimes of IB analysis part \emph{vs} flow number for the two approaches to check this intuition (right graph of Figure \ref{fig:compared_runtimes}).
The result is very explicit: IB analysis of BATA is faster than G-BATA. 
For instance, on 48-flow configurations, BATA is on average 5.7 times faster than G-BATA. 

In an attempt to be more platform-independent, we have used other metrics than runtimes to estimate the complexity of analyses.
To do so, we counted the number of calls of relevant functions.
For the end-to-end delay bounds computations, we counted the total number of calls to the function \verb+endToEndServiceCurve()+ in Algorithm \ref{alg:end_to_end_service_computation}, which is used in both approaches.
For the IB analysis part, the two approaches are significantly different; thus, we counted the number of iterations of the \emph{while} loop for BATA and the number of calls to the function \verb+addVertex()+ for G-BATA when this function creates a new vertex. 
The number of calls to \verb+addVertex()+ in G-BATA is roughly the equivalent of the number of \emph{while} iterations of BATA.

We gathered the results in Figure \ref{fig:compared_call_nb}. We plotted two graphs: one for the service curve computation (left), the other one for the IB analysis (right).
The results match what the runtime graphs showed: G-BATA is way faster on the end-to-end service curve computation, while BATA is faster on IB analysis.
More precisely, for the total analysis of 48-flow configurations, BATA performs on average 1883 times as many calls to \verb+endToEndServiceCurve()+ as G-BATA does. 
For the IB analysis, G-BATA performs on average 1.5 times as many IB analysis iterations as BATA does. 

\begin{figure*}[!htb]
\centering
\includegraphics[scale=0.6]{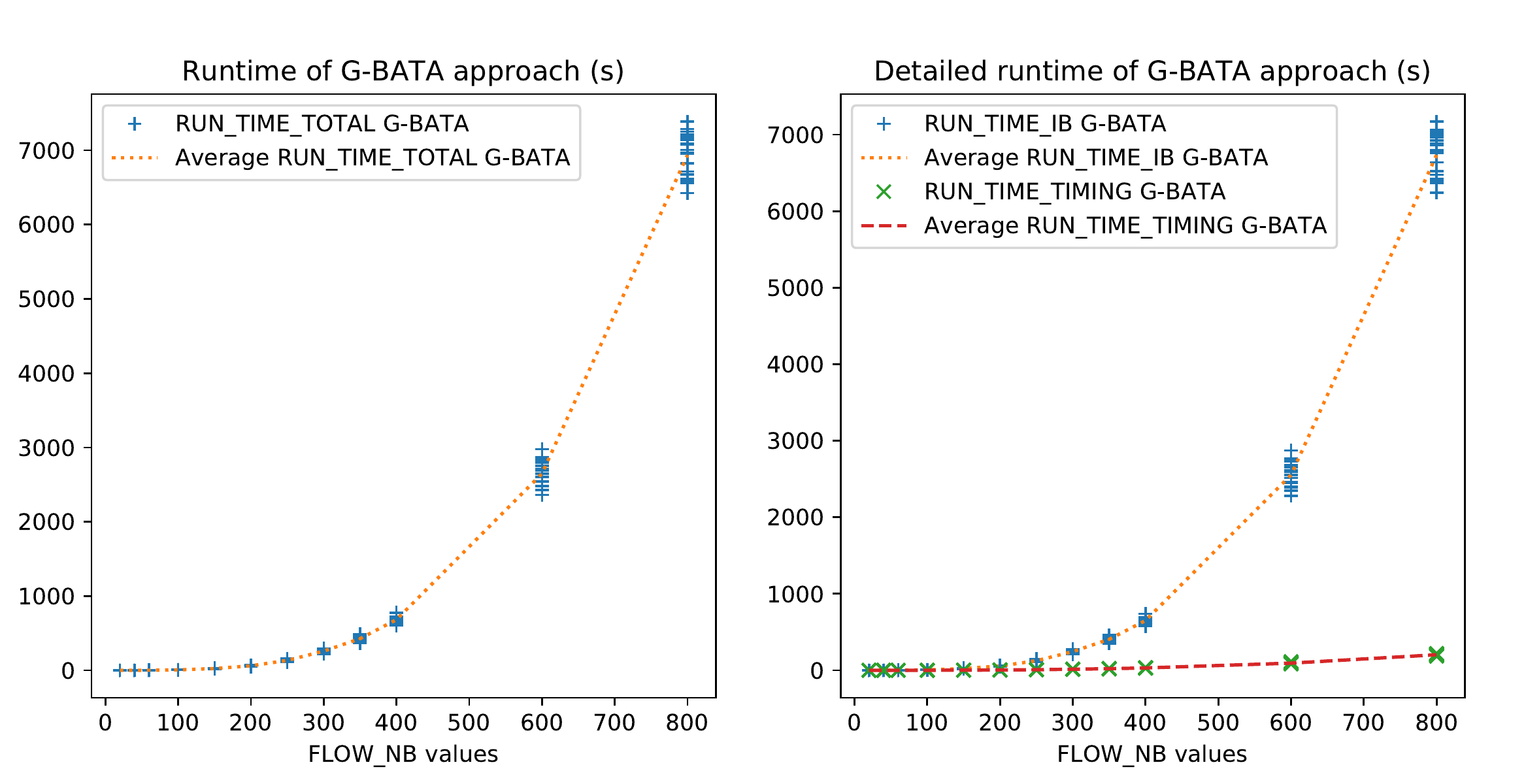}
\caption[Scalability of G-BATA on large flow sets]{Studying the scalability of G-BATA on large flow sets. RUN\_TIME\_IB denotes the duration of the IB analysis, while RUN\_TIME\_TIMING denotes the duration of the service curve computation.}
\label{fig:scalability_study_many_flows}
\end{figure*}

We then performed additional experiments on randomly generated configurations for G-BATA approach, on a $8\times 8$ NoC, with a number of flows from 20 to 800, to study how well the new method scales on large flow sets.
As before, we perform the analysis and measure total runtime, runtime of the IB analysis and runtime of the service curve computation. We plot the results on Figure \ref{fig:scalability_study_many_flows}.
What comes out of this additional study is that G-BATA analysis scales well: without parallelization, on a laptop powered by an Intel core i7 processor, computing end-to-end delay bounds for each of the 800 flows takes around 7200 seconds in the worst case (2 hours), \ie{} around 9 seconds per flow, as shown on the left graph of Figure \ref{fig:scalability_study_many_flows}.
Moreover, the IB analysis runtime is the more computationally expensive phase: for the 800-flow configurations, it represents on average 97.1\% of the total runtime, as illustrated on the right graph of Figure \ref{fig:scalability_study_many_flows}. 

\textbf{Key points:} G-BATA approach scales way better than BATA approach. The difference is especially visible for flow sets of 32 and 48 flows, where the average runtime of the total analysis for BATA is 10 to 100 times higher than G-BATA. For bigger configurations, we have not been able to get much comparative information as running one analysis with BATA takes more than two hours.
Moreover, G-BATA approach performs well on heavy configurations (600 and 800 flows) with an average total runtime of 2647 and 6935 seconds, respectively.
Finally, we notice that depending on the approach, the more computationally expensive step is either the indirect blocking analysis (G-BATA) or the service curve computation (BATA). For the latter case, it is what limits BATA approach scalability for large flow sets. \\

\begin{figure}[ht]
    \centering
    \resizebox{0.7\columnwidth}{!}{\input{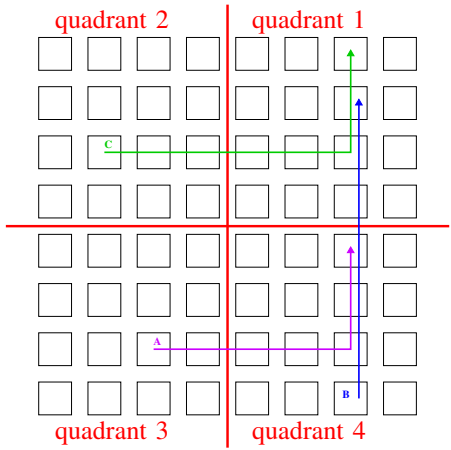}}
    \caption{Quadrants of the NoC and illustration of flows from families A, B and C}
    \label{fig:noc_quadrants}
\end{figure}

From these illustrated results, we can notice that for a given number of flows, runtimes can vary significantly from one configuration to another. 
For instance, on the left graph of Figure \ref{fig:compared_runtimes}, for 48-flow configurations, runtimes differ by up to 57\% and up to 99\% for G-BATA and BATA, respectively. 
Hence, the configuration complexity seems to not only depend on the number of flows, but also on at least another hidden parameter. 

In an attempt to better understand what are the configuration parameters impacting the approach complexity, we define two congestion indexes.
\begin{Definition}
    Given a configuration $\mathcal{F}$ and a \foi{} $f$, the direct blocking index (DB index) of $f$, denoted $I_{DB}(f)$, is the number of flows in the direct blocking set of $f$:
    \[ I_{DB}(f) = | \DB_f | \]
\label{def:db_index}
\end{Definition}

\begin{Definition}
    Given a configuration $\mathcal{F}$ and a \foi{} $f$, the indirect blocking index (IB index) of $f$, denoted $I_{IB}(f)$, is the number of \{flow index, subpath\} pairs in the indirect blocking set of $f$:
    \[ I_{IB}(f) = | \IB_f | \]
\label{def:ib_index}
\end{Definition}

The value of one such index is specific to one flow. Hence, to quantify how complex a configuration is, we introduce the following average indexes:
\begin{itemize}
    \item $|\mathcal{F}|$, the number of flows of the configuration;
    \medskip
    \item $\overline{I_{IB}} = \frac{1}{|\mathcal{F}|} \sum\limits_{f \in \mathcal{F}} I_{IB}(f)$, the average IB index of flow set $\mathcal{F}$;
    \item $\overline{I_{DB}} = \frac{1}{|\mathcal{F}|} \sum\limits_{f \in \mathcal{F}} I_{DB}(f)$, the average DB index of flow set $\mathcal{F}$.
\end{itemize}

To evaluate the impact of these introduced indicators on the runtime of BATA and G-BATA, 
we randomly generated another series of 4-, 8-, 16- and 32-flow configurations (20 configurations per number of flows), but this time following a different paradigm.
We split the NoC into 4 quadrants (Figure \ref{fig:noc_quadrants}).
Then, we randomly choose flows according to 3 different sets, A, B and C:
\begin{itemize}
  \item flows in A have their source in the 3$^{\textrm{rd}}$ quadrant and their destination in the 4$^{\textrm{th}}$ quadrant;
  \item flows in B have their source in the 4$^{\textrm{th}}$ quadrant and their destination in the 1$^{\textrm{st}}$ quadrant;
  \item flows in C have their source in the 2$^{\textrm{nd}}$ quadrant and their destination in the 1$^{\textrm{st}}$ quadrant.
\end{itemize}
It is worth noticing that these communication patterns favor direct and indirect blocking, which impact the introduced direct and indirect blocking indexes. 

\begin{figure*}[!htb]
\centering
\includegraphics[scale=0.5]{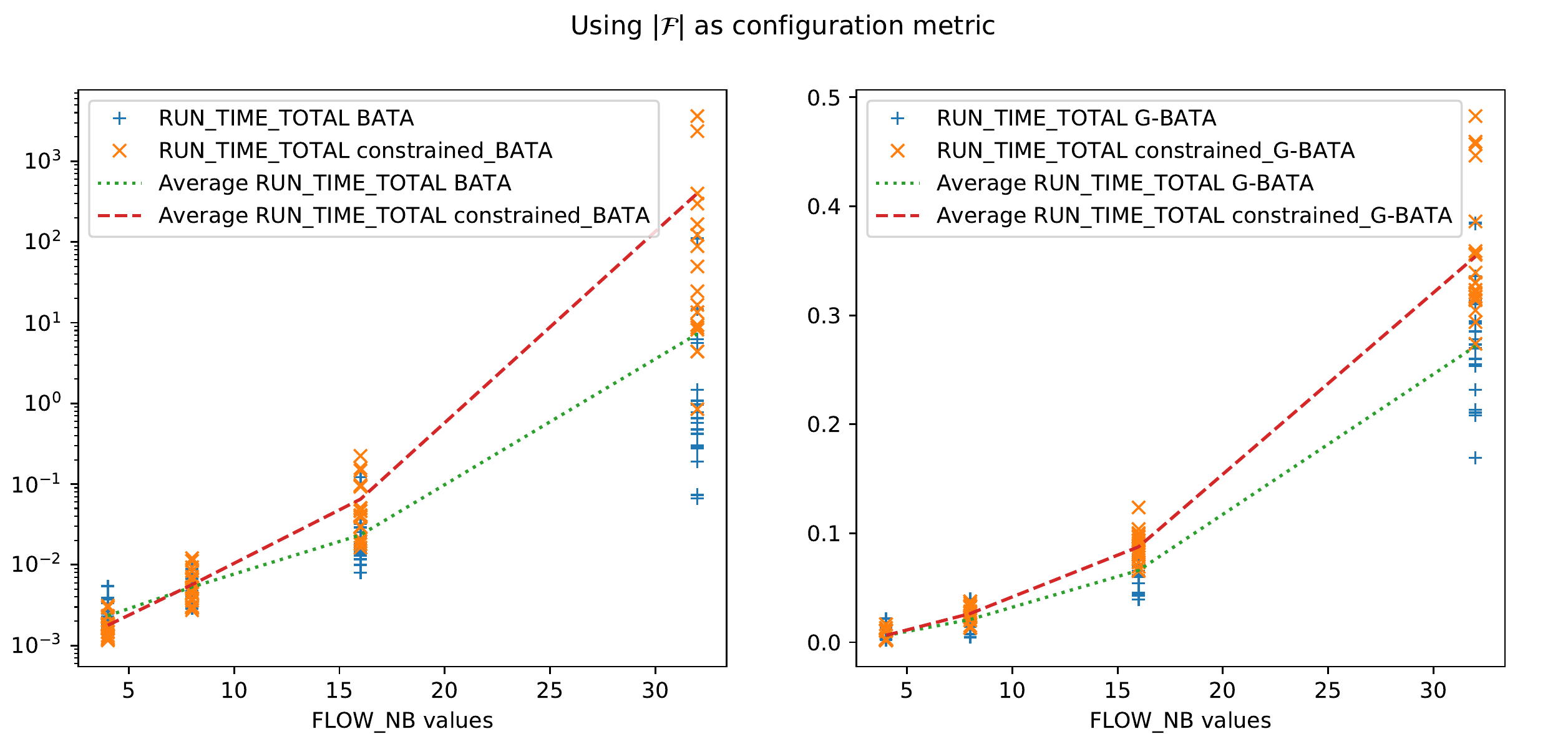}
\caption[Runtimes \emph{vs} flow number]{Runtimes \emph{vs} flow number for both configuration types for BATA (left) and G-BATA (right)}
\label{fig:runtimes_vs_flow_number}
\end{figure*}

We perform the same analysis as before and compare the results we get for both approaches, on these constrained configurations (referred to as ``constrained'') and the previous 4-, 8-, 16- and 32-flow configurations.

We first plot total runtime as a function of flow number, and the average curve (Figure \ref{fig:runtimes_vs_flow_number}). 
We notice that for both G-BATA and BATA approaches, there is a noticeable difference between the constrained and the uniformly distributed configurations.
For a given number of flows, constrained sets generally require greater runtimes than the previous sets.
We did not include the plots of other runtimes (IB analysis and service curve computation) \emph{vs} flow number for the two configuration types, but they exhibit the same trend as total runtimes.

\begin{figure*}[!htb]
\centering
\includegraphics[scale=0.5]{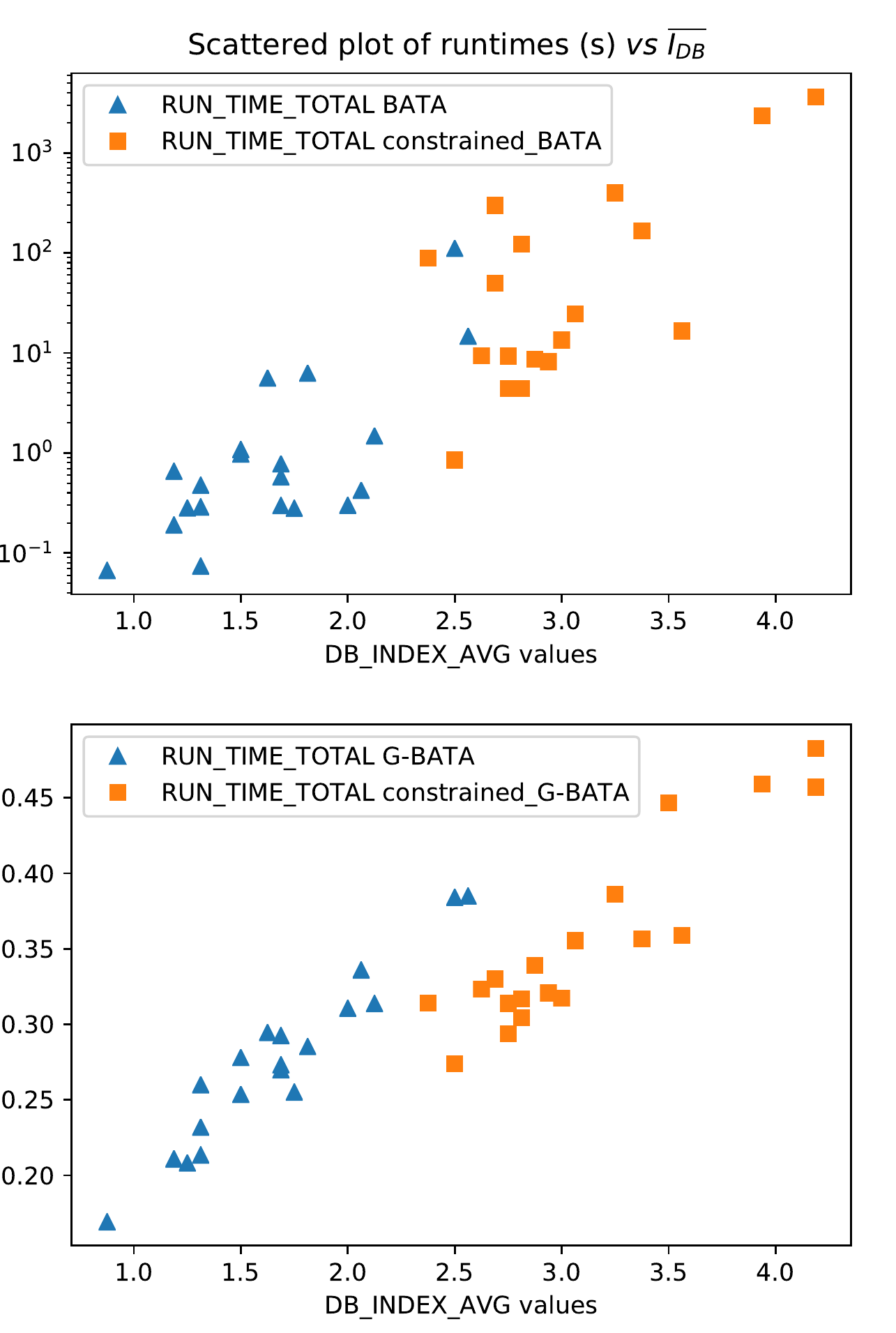}\includegraphics[scale=0.5]{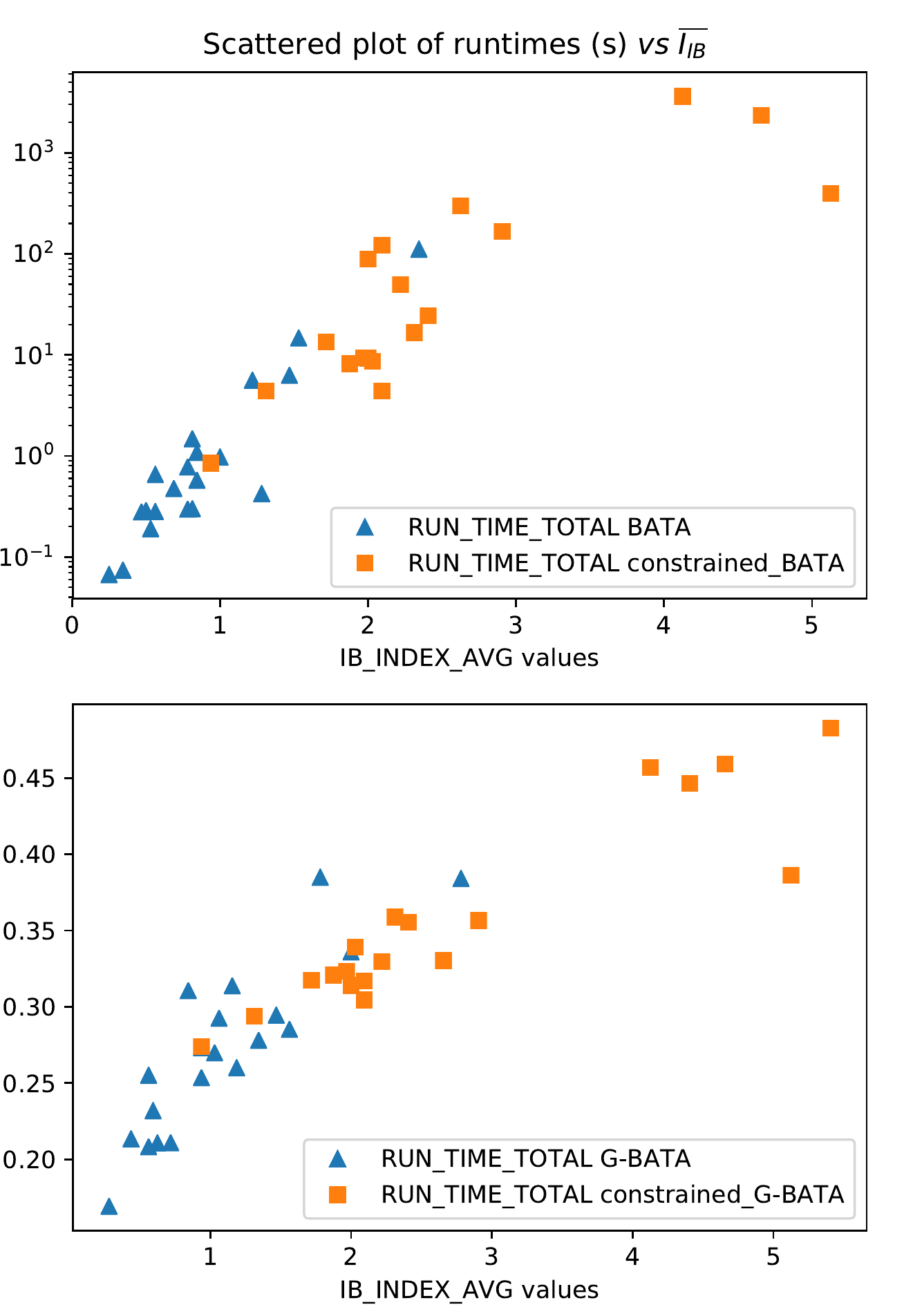}
\caption[Studying the correlation between average DB index (resp. average IB index) and total runtime for 32-flow configurations]{Studying the correlation between average DB index (resp. average IB index) and total runtime for 32-flow configurations, for BATA (top graphs) and G-BATA (bottom graphs)}
\label{fig:scalability_study_congestion_index}
\end{figure*}

Hence, to better understand the correlation between the runtime and the congestion pattern, we focus on the 32-flow configurations and we plot, for both approaches, all points $(x,y)$ where:
\begin{itemize}
\item $x$ is the average DB index (resp. IB index) of the configuration, $\overline{I_{DB}}$ (resp. $\overline{I_{IB}}$);
\item $y$ is the total analysis runtime.
\end{itemize}
The results are gathered in Figure \ref{fig:scalability_study_congestion_index}.
For both approaches, we notice that the runtime tend to increase with the average congestion index (direct or indirect). 
We conclude that a higher average congestion index (direct or indirect) tends to characterize configurations that require a higher computation time.

Moreover, the average IB index does not bring more insights than the average DB index on how computationally expensive the analysis of a configuration may be.
So, given that it is computationally more expensive to compute the average IB index than the average DB index, especially for G-BATA approach, 
we conclude that average DB index is a good configuration indicator to quantify the complexity of a configuration in addition to the number of flows.

\textbf{Key points:} Although there is a correlation between the number of flows of one set and the runtime needed to perform its analysis, we find that it is not sufficient to characterise how long the timing analysis may take. 
In that respect, we propose two configuration indicators to refine the quantitative aspect of the complexity of a flow set: the average DB and IB indexes. 
We show that both are adequate complementary configuration parameters. 
Nonetheless, the average IB index is computationally more expensive while not bringing much more information. 
Hence, the DB index and the size of the flow set are considered as sufficient to characterize a configuration complexity. \\

\subsection{Sensitivity Analysis}
\label{subsec:sensitivity_gbata}

In this section, we study the impact of different parameters on the end-to-end delay bounds yielded by G-BATA. 
For the sensitivity analysis, we will analyze the end-to-end delay bounds when varying the following parameters: 
\begin{itemize}
    \item buffer size for values 1, 2, 3, 4, 6, 8, 12, 16, 32, 48, 64 flits;
    \item total packet length (including header) for values 2, 4, 8, 16, 64, 96, 128 flits;
    \item flow rate for values between 1\% and 40\% of the total link capacity (so that the total utilization rate on any link remains below 100\%).
\end{itemize}

\begin{figure}[ht]
\rule{0pt}{12pt}\\
\centering
\begin{minipage}{0.4\textwidth}
\renewcommand{\arraystretch}{1}
\begin{tabular}{@{}lll@{}}
\toprule
Flow & Source core & Destination core \\
\midrule
1    & (0, 5)      & (5, 4) \\
2    & (1, 5)      & (2, 3) \\
3    & (2, 5)      & (3, 2) \\
4    & (3, 5)      & (4, 3) \\
5    & (5, 5)      & (5, 1) \\
6    & (2, 4)      & (2, 1) \\
7    & (2, 2)      & (2, 0) \\
8    & (3, 4)      & (3, 1) \\
9    & (3, 3)      & (3, 0) \\
10   & (4, 4)      & (4, 1) \\
11   & (4, 2)      & (4, 0) \\
12   & (5, 2)      & (5, 0) \\
\bottomrule
\end{tabular}
\renewcommand{\arraystretch}{1.5}
\end{minipage}\\
\rule{0pt}{12pt}\\
\begin{minipage}{0.4\textwidth}
\resizebox{0.7\columnwidth}{!}{\input{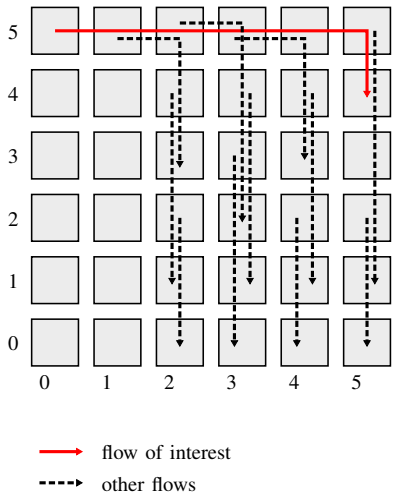}}
\end{minipage}
\caption{Flow configuration on a 6$\times$6 mesh NoC}
\label{fig:sensitivity_configuration}
\end{figure}

To achieve this aim, we consider the configuration described on Figure \ref{fig:sensitivity_configuration}. 
This configuration remains quite simple but exhibits sophisticated indirect blocking patterns. We assume periodic flows with no jitter having the same period and packet length, and consider the following parameters: 
\begin{itemize}
    \item each router can handle one flit per cycle and it takes one cycle for one flit to be forwarded from the input of a router to the input of the next router, \ie{}, for any node $r$, $T^r=1$ cycle and $R^r=1$ flit/cycle;
    \item all the flows are mapped on the same VC;
    \item our flow of interest is flow 1.
\end{itemize}

To better highlight the impact of the various parameters on G-BATA in reference to BATA, we display the results of G-BATA along with the existing results obtained with BATA. 

\begin{figure*}[ht]
    \centering
    \vspace{-10pt}
\rule{-1cm}{0cm}\subcaptionbox{Constant rate\label{subfig:constant_rate}}{\includegraphics[scale=0.5,trim=0mm 4mm 0mm 0mm]{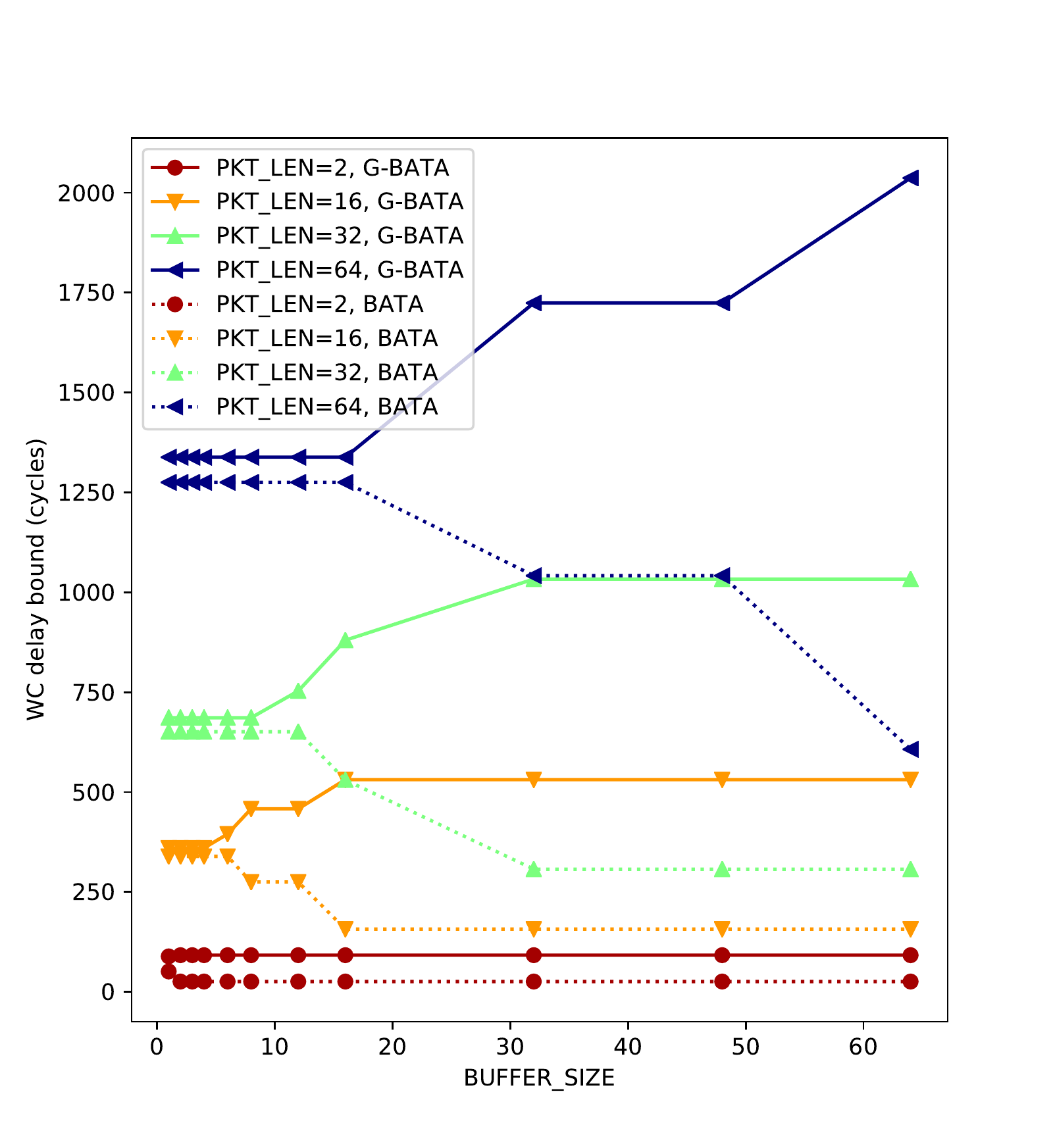}}\subcaptionbox{Constant packet length\label{subfig:contant_pktlen}}{\includegraphics[scale=0.5]{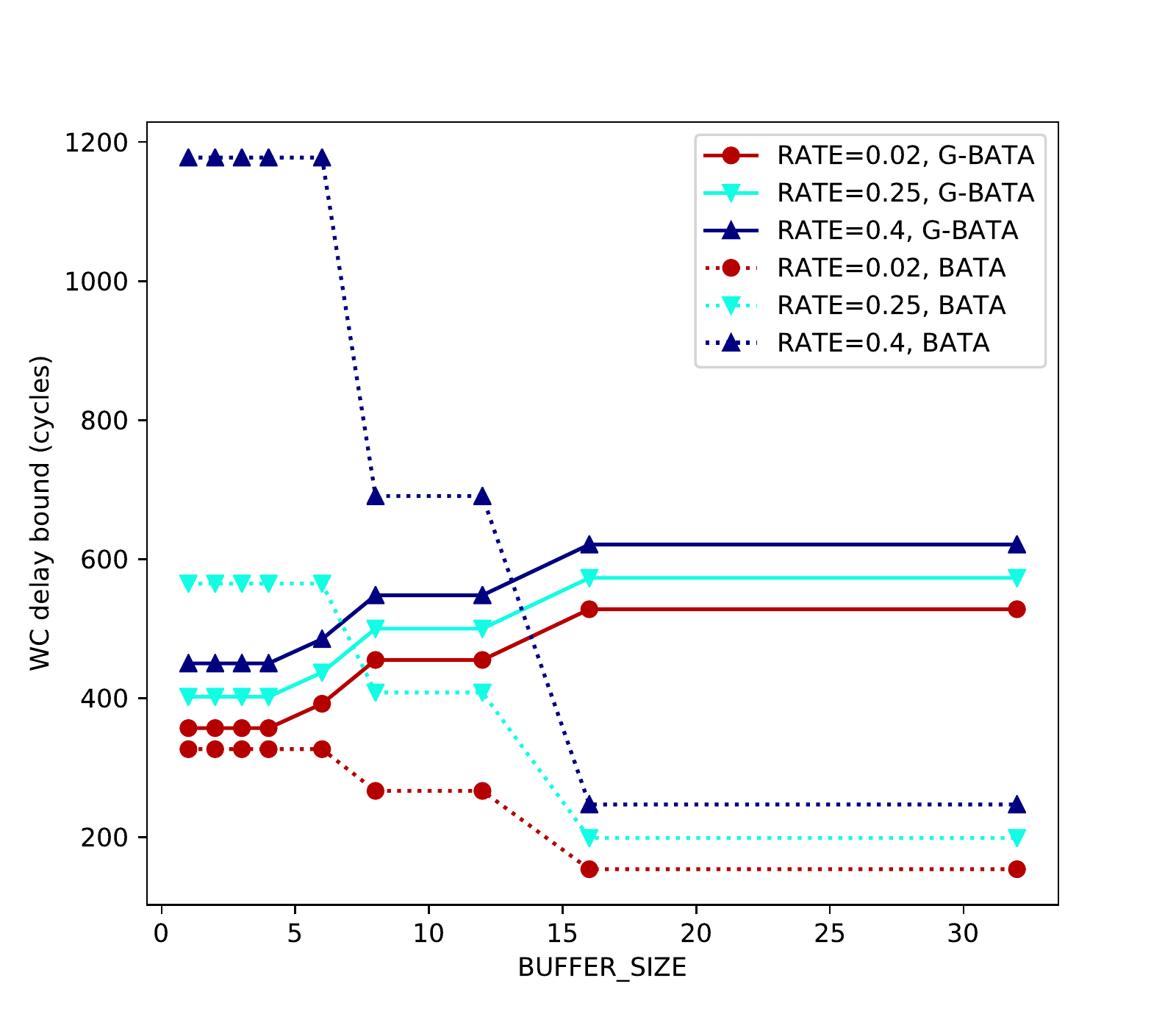}}
    \caption{Compared buffer size impact on end-to-end delay bounds}
    \label{fig:buffer_size_impact}
\end{figure*}

Figure \ref{fig:buffer_size_impact} illustrates the end-to-end delay bounds of the \foi{} when varying buffer size. For the left graph, we keep each flow rate constant at 4\% of the total bandwidth; whereas for the right graph, we keep each flow packet length at 16 flits. 

First, on both graphs, we notice an opposite trend between G-BATA and BATA approaches. 
The former predicts that delay bounds increase when buffer size increases, whereas the latter predicts that delay bounds decrease. 
This is mainly due to the variation of the spread index of flows and its impact on each approach. 

For BATA, this generally makes the IB set smaller: as this approach does not consider CPQ, reducing the length of a subpath reduces the possibility that this subpath intersects with the paths of other flows. 
Consequently, the derived IB latency tends to decrease, as well as the end-to-end delay bound. 

For G-BATA approach, however, the interference graph takes CPQ into account, and in that respect, the number of consecutive packets is not bounded. 
Therefore, reducing the size of the subpaths increases their number. 
The extracted IB set thus contains more subpaths of smaller size. 
Consequently, there are more terms in the indirect blocking delay sum (Equation \ref{eqn:t_ib_graph}), which may increase the end-to-end delay bound. \\

Second, we notice that with both approaches, the end-to-end delay bounds increase with the packet length and rate. 
Moreover, we observe that past a certain value of buffer size, the end-to-end delay bounds remain constant. 
This corresponds to the IB set remaining constant once buffers are large enough to hold one packet (spread index of 1 for all flows).

Finally, on the right graph, we notice that BATA is more sensitive to rate than G-BATA:
for buffer sizes below 6 flits, BATA predicts delay bounds between 327 and 1178 cycles, while G-BATA gives delay bounds between 357 and 486 cycles. \\

\textbf{Key points:} Although increasing buffer size may improve end-to-end delay bounds when no CPQ happens (under BATA), we find that it does not impact favorably the end-to-end delay bound when CPQ can occur and the number of consecutive packets queueing is not limited. 
Moreover, G-BATA is less sensitive to rate variations than BATA for small buffer sizes. \\

\begin{figure*}[ht]
\centering
\rule{-1cm}{0cm}\includegraphics[scale=0.5]{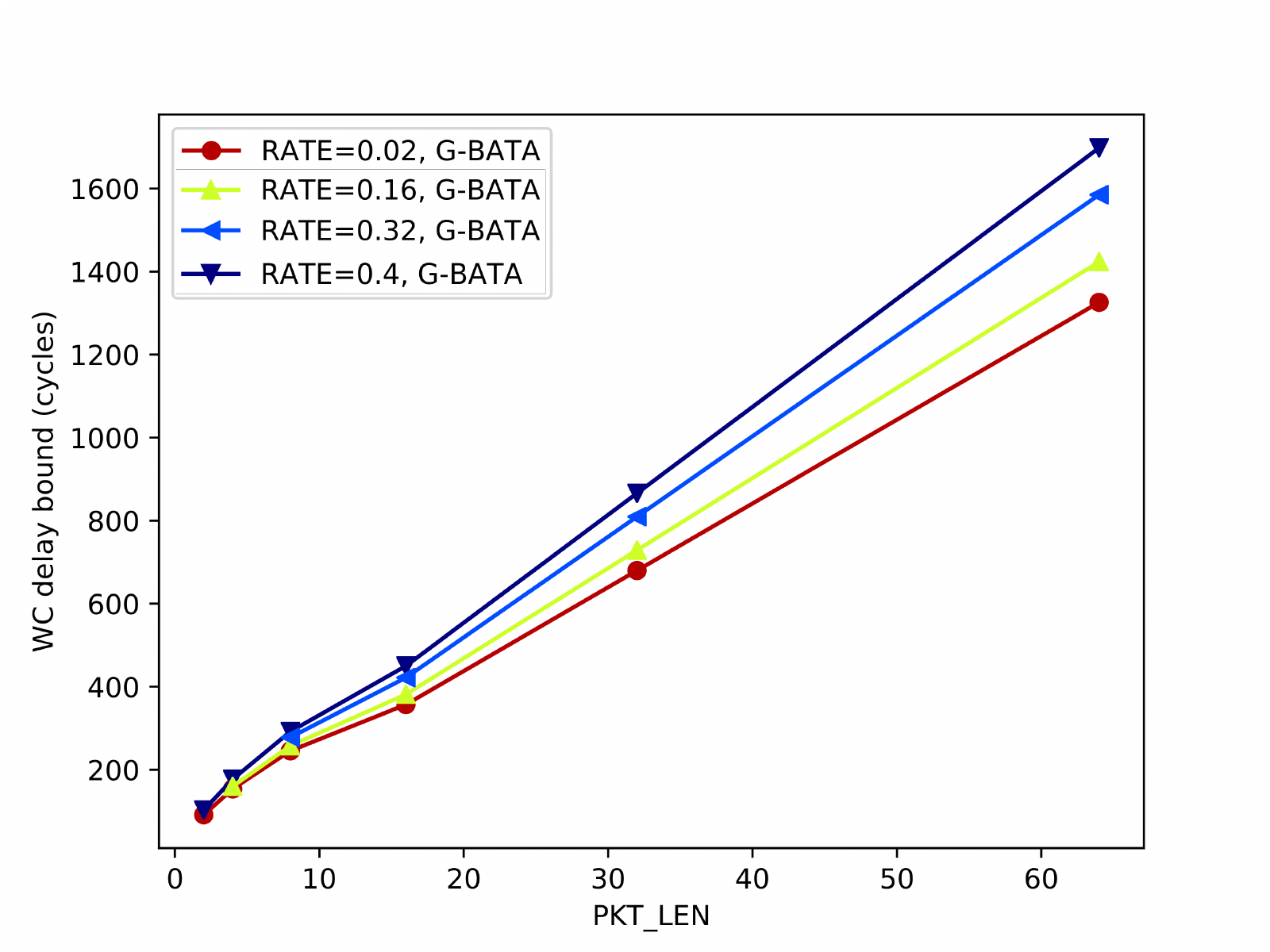}\includegraphics[scale=0.5]{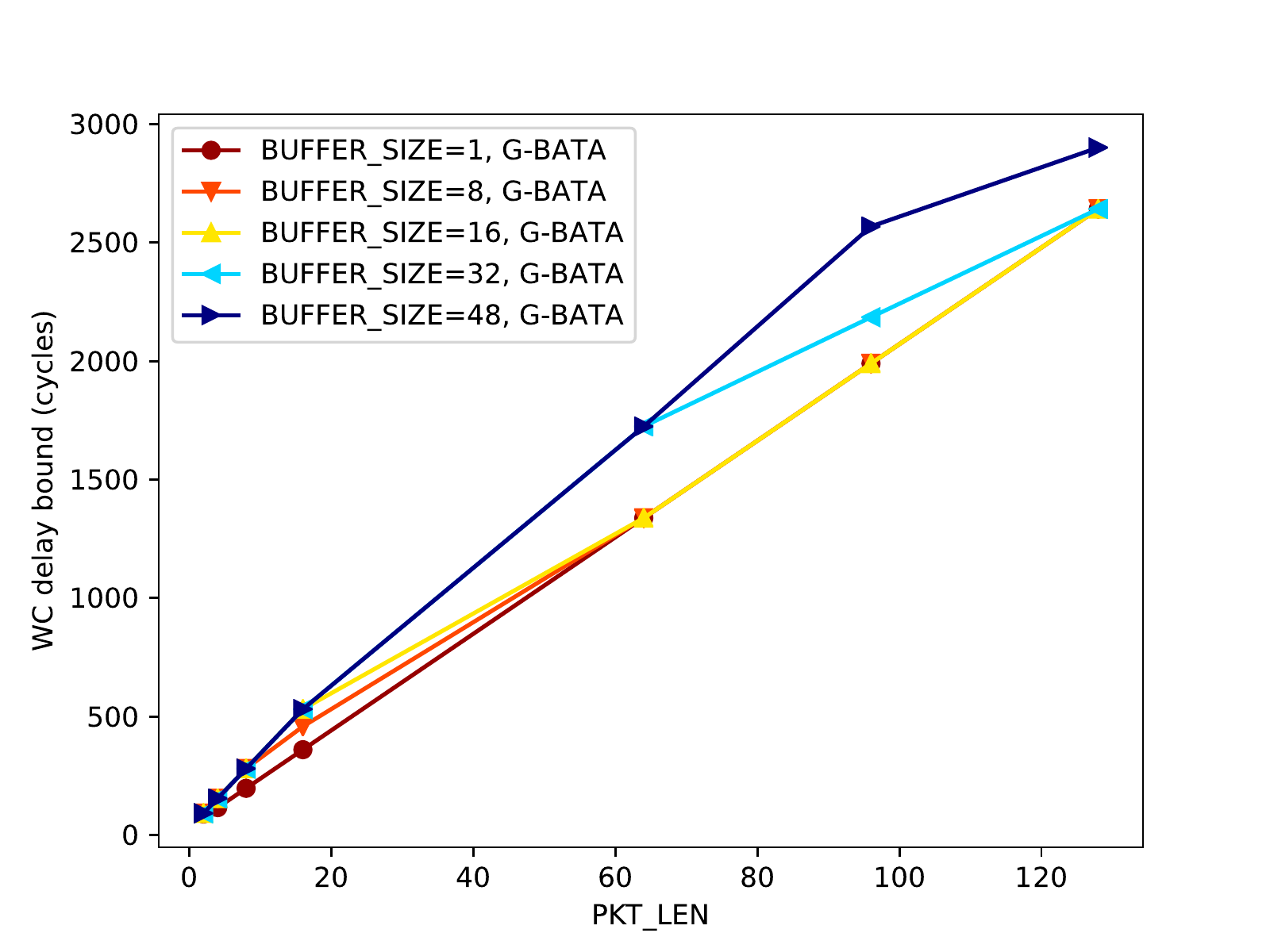}
    \caption{Packet length impact on G-BATA end-to-end delay bounds}
    \label{fig:payload_impact_graph}
\end{figure*}

\begin{figure*}[ht]
    \centering
\rule{-1cm}{0cm}\includegraphics[scale=0.5]{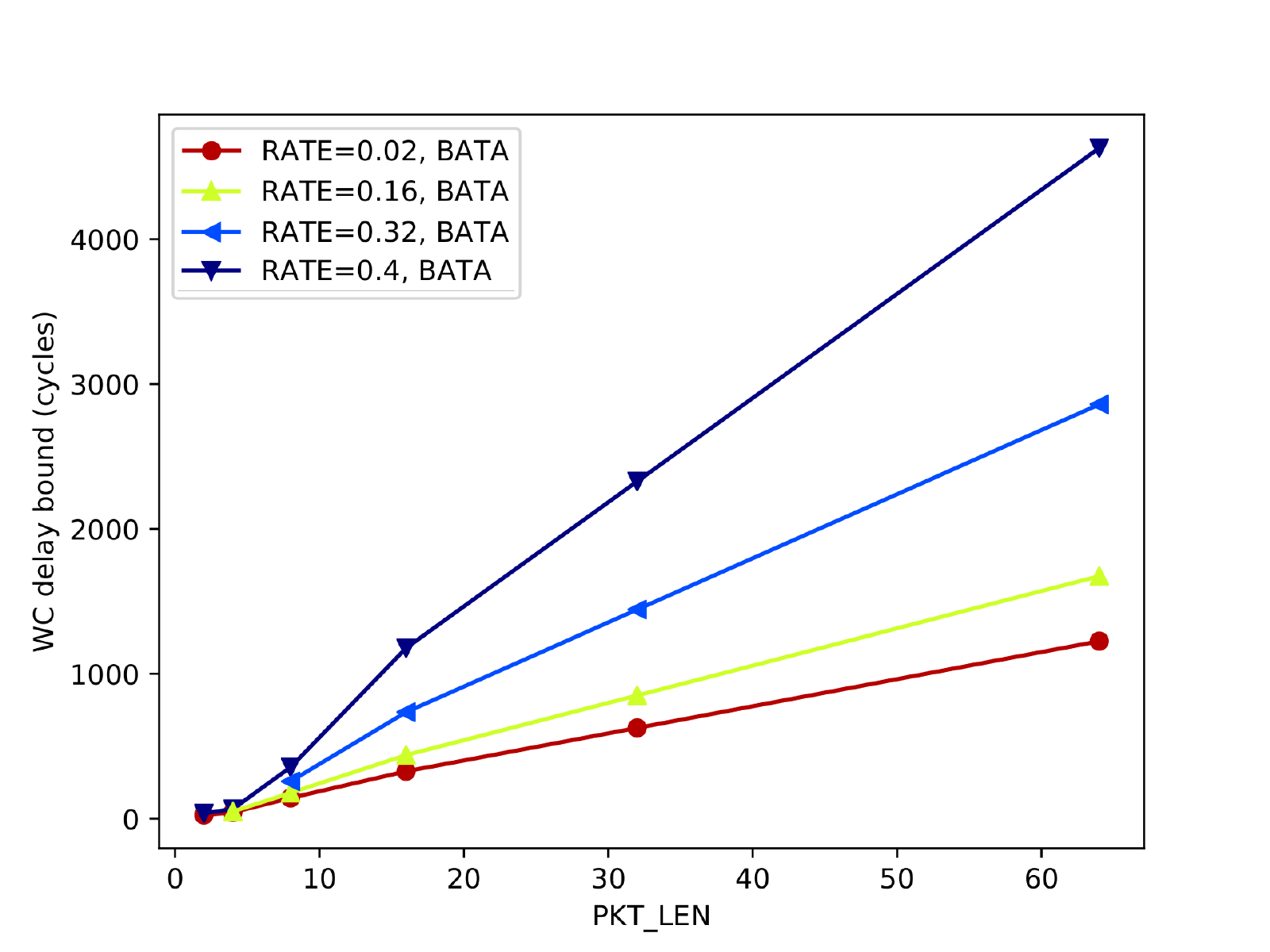}\includegraphics[scale=0.5]{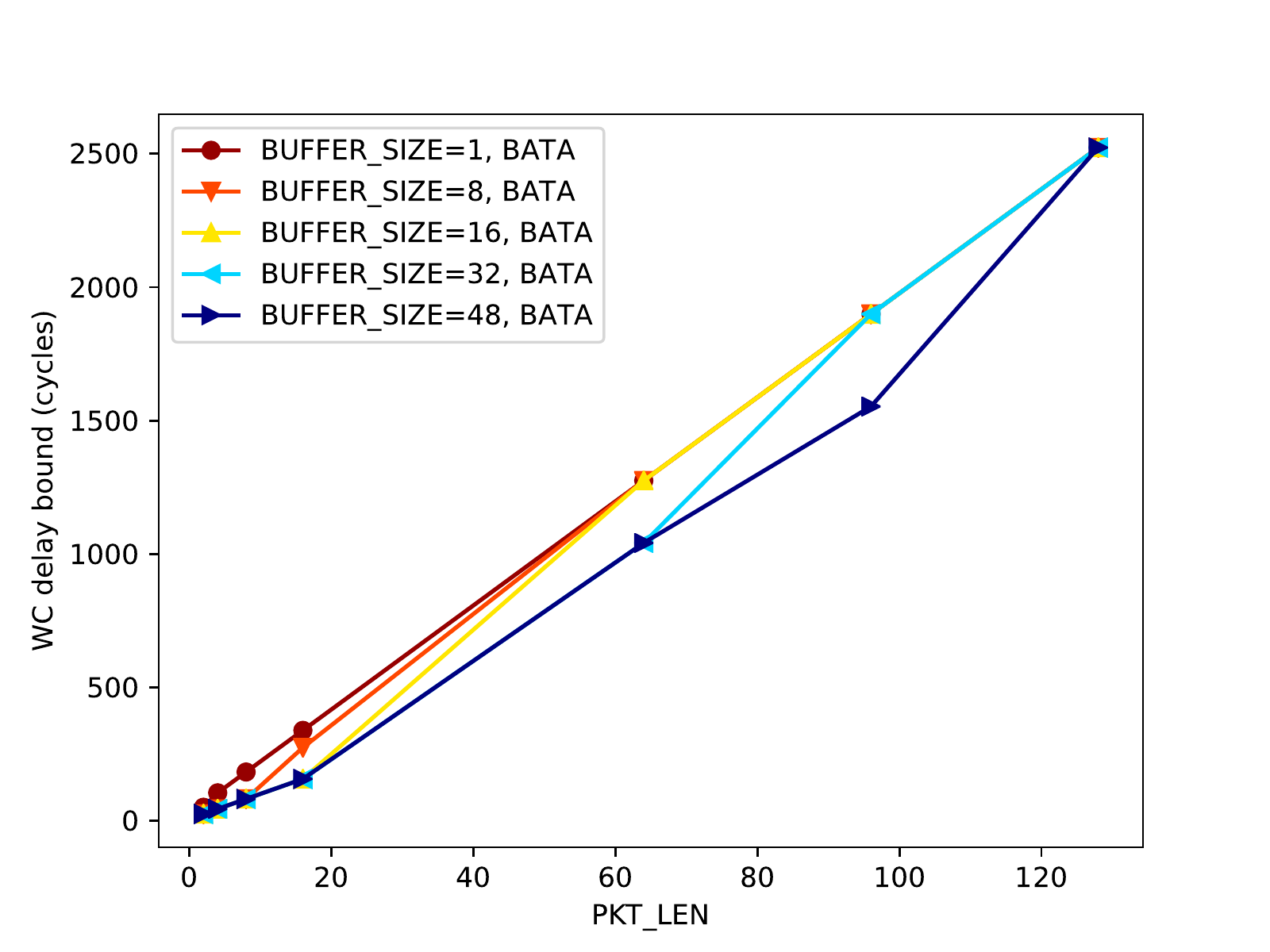}
    \caption{Packet length impact on BATA end-to-end delay bounds}
    \label{fig:payload_impact_rtas18}
\end{figure*}

Next, we focus on the packet length impact on the end-to-end delay bound for G-BATA and BATA, as illustrated on Figure \ref{fig:payload_impact_graph} and \ref{fig:payload_impact_rtas18}, respectively. 
For clarity reasons, we plotted separate graphs for the two approaches. 
On each figure, the left graphs present results when the buffer size is constant (4 flits) and the right ones when the rate of each flow is constant (4\% of the link capacity). 


The first observation we can make from all graphs is that the delay bounds evolve in an almost linear manner with the packet length. 
For instance, on the right G-BATA graph, with 8 flits of buffer size and packet length equal to 64, 96 and 128 flits, the ratios of packet length and end-to-end delay bound are 20.9, 20.7 and 20.6, respectively. 

Still on the same right graph, we observe further interesting aspects:
\begin{itemize}
\item At a given packet length, the buffer size has a limited impact on the end-to-end delay bounds. For instance, for a packet length of 64 flits, the delay bounds increase with less than 30\% when the buffer size increases with 480\%;
\item For packet lengths that are significantly larger than buffer size, the delay bound remains constant regardless of the buffer size, \eg{}, it is the case for a packet length of 128 flits.
\end{itemize}
Similar observations can be made for BATA approach. 

However, looking at the left graphs for BATA and G-BATA, we notice that BATA is more sensitive to rate variations than G-BATA: for a packet of 64 flits, when the rate increases from 2\% to 40\%, the end-to-end delay bound yielded by BATA increases from 1226 cycles to 4630 cycles (+278\%) while the delay bound predicted by G-BATA increases only from 1326 cycles to 1698 cycles (+28\%). \\


\textbf{Key points:} at a given rate and packet length, we observe that buffer size has a limited impact on the end-to-end delay bound, and this observation is valid for both G-BATA and BATA approaches. 
We also notice that the evolution of the delay bound with the packet length follows an almost linear trend, for both approaches as well. 
Finally, we further confirm that BATA is more sensitive to rate variations than G-BATA, especially for large packet lengths. \\

\begin{figure*}[ht]
\centering
\rule{-1cm}{0cm}\includegraphics[scale=0.5]{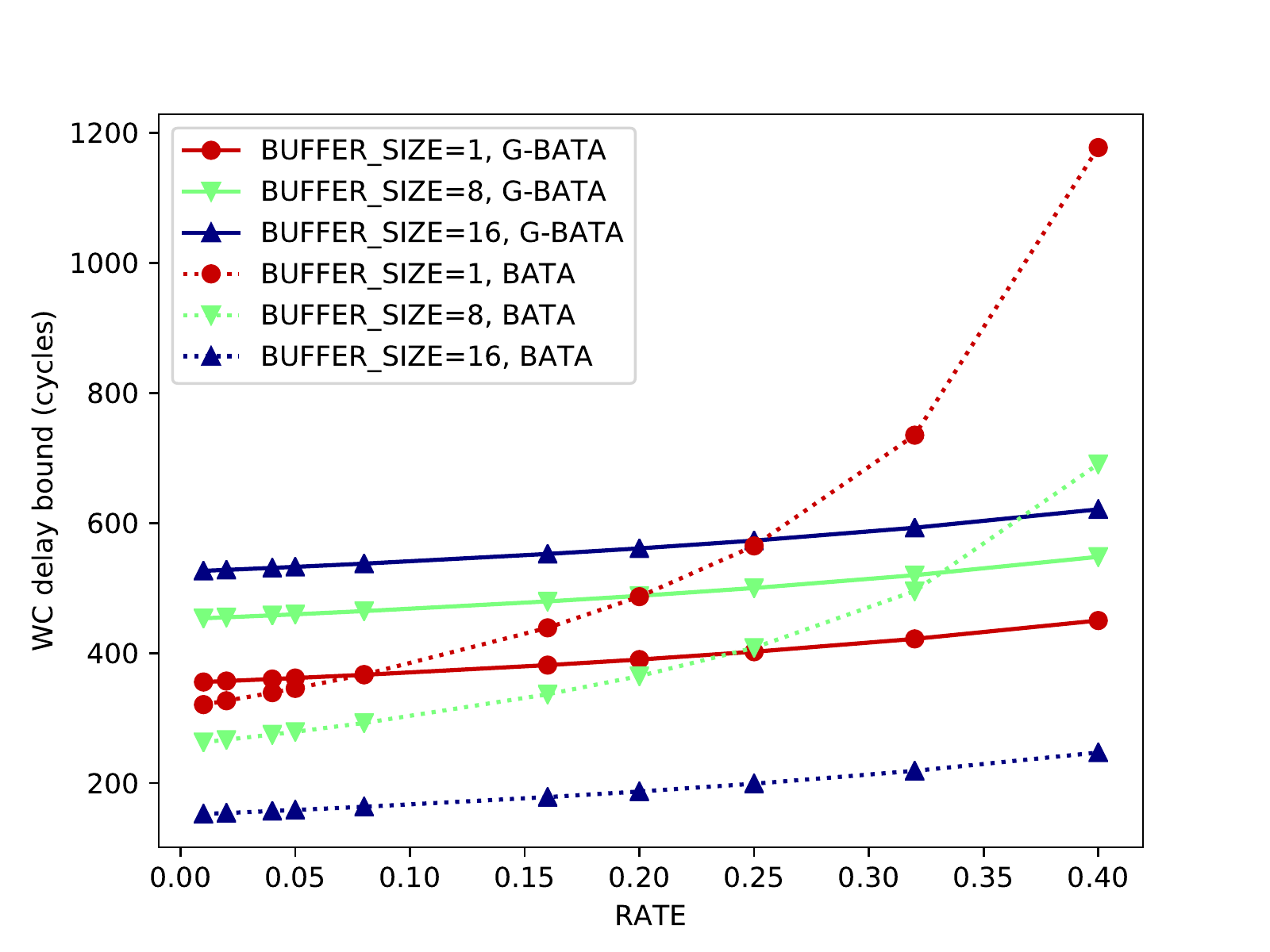}\includegraphics[scale=0.5]{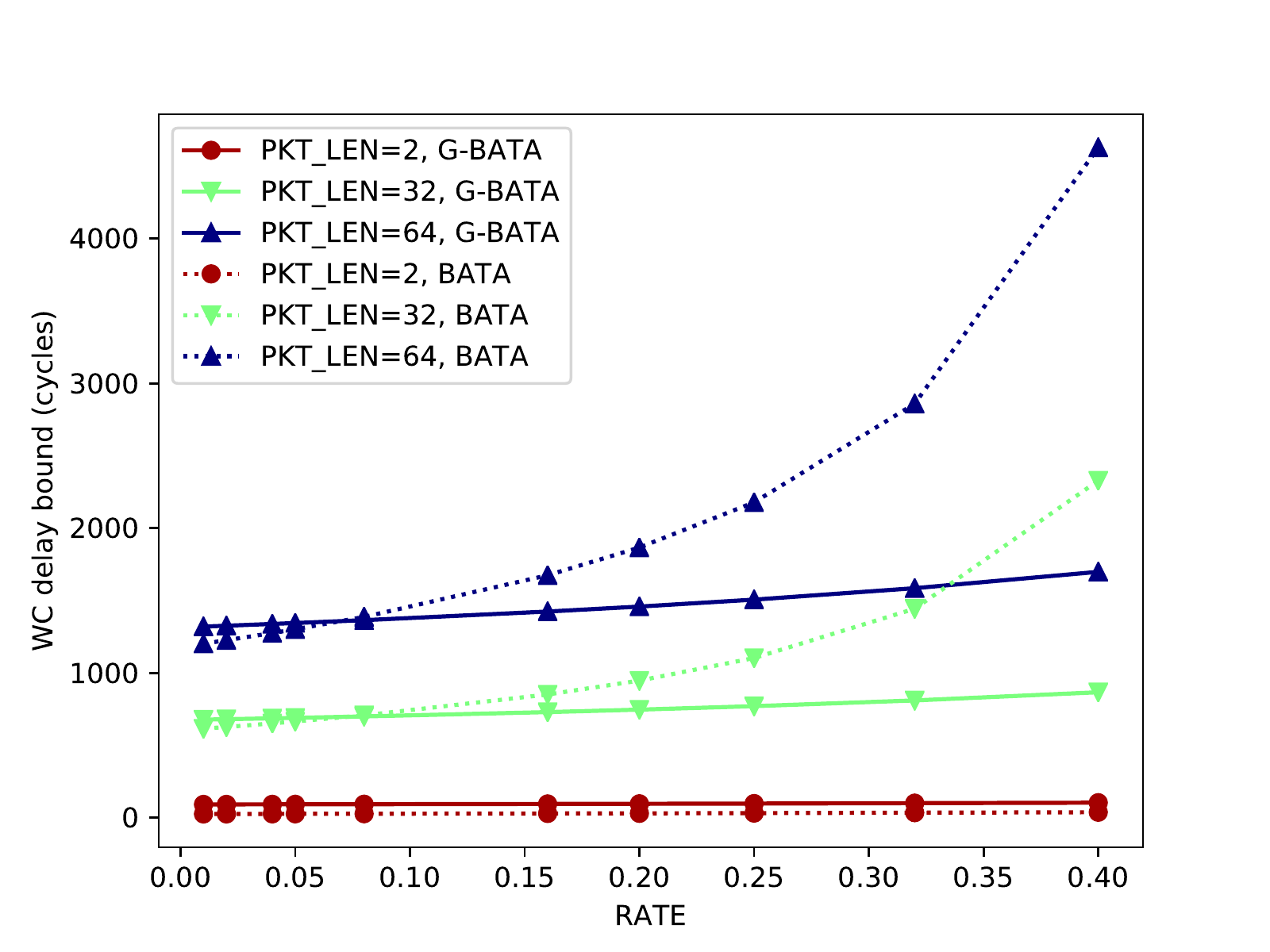}
\caption{Compared flow rate impact on end-to-end delay bounds}
\label{fig:rate_impact_gbata}
\end{figure*}

We now focus on the impact of the flow rate on end-to-end delay bounds (Figure \ref{fig:rate_impact_gbata}). 
The left graph represents the evolution of delay bounds when packet length is fixed (16 flits) for different values of buffer size, and the right graph shows the evolution of delay bounds with a fixed buffer size (4 flits) and values of packet length from 2 to 64 flits. 
As expected, with both approaches, the end-to-end delay bound increases with the rate. 
What is more interesting is that delay bounds with G-BATA approach increase much less rapidly than with BATA approach for buffers of 1 and 8 flits: at a 40\% flow rate, BATA gives bounds that are 26\% to 162\% greater than bounds given by G-BATA approach (left graph on Figure \ref{fig:rate_impact_gbata}). 
Therefore, we can confirm one more time that BATA is more sensitive to rate variations than G-BATA. 

Although there is generally no strict order between the bounds given by the two approaches, for instance for $B=8$ flits, we can notice a trend regarding the relative position of the bounds: 
BATA predicts smaller bounds than G-BATA for large buffer sizes and small rates, and the trend is opposite for small buffer sizes, especially as the rate increases.
When the rate of flow $\rho$ increases with all other parameters constant, the propagated burst of an arrival curve increases by $\rho \cdot T$ per node with a service curve latency of $T$.
Results obtained with BATA are especially impacted by this burst propagation since the burst is propagated at the beginning of the subpaths when computing $T_{\IB}$. 
This explains why BATA-predicted bounds increase faster than graph-predicted bounds when increasing the rate. 

\textbf{Key points:} Both approaches predict an increase of the end-to-end delay bound with the rate, however this increase is significantly different depending on the approach. 
Burst propagation at the beginning of subpaths in BATA approach leads to important bound increase when the flow rate is high. 
For instance, the computed bounds are up to 275\% higher with BATA than with G-BATA at 40\% flow rate. \\


\subsection{Tightness Analysis}
\label{subsec:tightness_gbata}

\renewcommand{\arraystretch}{1}
\begin{table*}
\centering
\begin{tabular}[ht]{@{}rrrrrrr@{}} 
\toprule
      & \multicolumn{6}{l}{$B=4$} \\
      & \multicolumn{2}{l}{rate = 8\%} & \multicolumn{2}{l}{rate = 32\%} & & \\
Flow  & \multicolumn{1}{l}{G-BATA} & \multicolumn{1}{l}{BATA} & \multicolumn{1}{l}{G-BATA} & \multicolumn{1}{l}{BATA} & \textbf{$I_{DB}$} & \textbf{$I_{IB}$} \\ 
\midrule
 1    & 40\%     & 44\%     & 74\%     & 44\%     & 4  & 11 \\ 
 2    & 42\%     & 41\%     & 87\%     & 24\%     & 2  & 12 \\ 
 3    & 63\%     & 64\%     & 75\%     & 51\%     & 3  & 5  \\ 
 4    & 64\%     & 68\%     & 93\%     & 64\%     & 2  & 3  \\ 
 5    & 72\%     & 77\%     & 68\%     & 46\%     & 2  & 0  \\ 
 6    & 73\%     & 75\%     & 57\%     & 21\%     & 2  & 0  \\ 
 7    & 85\%     & 89\%     & 100\%    & 43\%     & 1  & 0  \\ 
 8    & 66\%     & 68\%     & 67\%     & 46\%     & 2  & 0  \\ 
 9    & 45\%     & 47\%     & 37\%     & 24\%     & 2  & 0  \\ 
10    & 84\%     & 88\%     & 88\%     & 70\%     & 2  & 0  \\ 
11    & 87\%     & 92\%     & 81\%     & 69\%     & 1  & 0  \\ 
12    & 85\%     & 89\%     & 91\%     & 66\%     & 1  & 0  \\ 
\midrule
avg   & 67.36\%  & 70.11\%  & 76.52\%  & 47.41\%  & -  & -  \\ 
min   & 40.20\%  & 40.61\%  & 36.51\%  & 20.77\%  & -  & -  \\ 
max   & 87.19\%  & 91.71\%  & 99.74\%  & 70.46\%  & -  & -  \\ 
\bottomrule
\end{tabular}
\begin{tabular}[ht]{@{}rrrrrrr@{}}
\toprule
      & \multicolumn{6}{l}{$B=16$} \\
      & \multicolumn{2}{l}{rate = 8\%} & \multicolumn{2}{l}{rate = 32\%} & & \\
Flow  & \multicolumn{1}{l}{G-BATA} & \multicolumn{1}{l}{BATA} & \multicolumn{1}{l}{G-BATA} & \multicolumn{1}{l}{BATA} & \textbf{$I_{DB}$} & \textbf{$I_{IB}$} \\ 
\midrule
 1    & 24\%     & 79\%     & 44\%      & 87\%     & 4  & 22 \\ 
 2    & 18\%     & 79\%     & 43\%      & 97\%     & 2  & 27 \\ 
 3    & 35\%     & 85\%     & 48\%      & 75\%     & 3  & 14 \\ 
 4    & 34\%     & 86\%     & 51\%      & 87\%     & 2  & 8  \\ 
 5    & 66\%     & 85\%     & 47\%      & 64\%     & 2  & 0  \\ 
 6    & 63\%     & 86\%     & 45\%      & 86\%     & 2  & 0  \\ 
 7    & 83\%     & 87\%     & 78\%      & 97\%     & 1  & 0  \\ 
 8    & 60\%     & 70\%     & 46\%      & 71\%     & 2  & 0  \\ 
 9    & 40\%     & 49\%     & 27\%      & 44\%     & 2  & 0  \\ 
10    & 80\%     & 88\%     & 76\%      & 88\%     & 2  & 0  \\ 
11    & 86\%     & 88\%     & 100\%     & 96\%     & 1  & 0  \\ 
12    & 83\%     & 87\%     & 76\%      & 88\%     & 1  & 0  \\ 
\midrule
avg   & 56.08\%  & 80.76\%  & 56.73\%   & 81.60\%  & -  & -  \\
min   & 18.07\%  & 48.94\%  & 26.96\%   & 43.83\%  & -  & -  \\
max   & 86.50\%  & 88.32\%  & 100.00\%  & 97.34\%  & -  & -  \\
\bottomrule
\end{tabular}\\
\medskip
\caption[Tightness summary for both approaches, buffer size 4 flits and 16 flits]{Tightness summary for both approaches, buffer size 4 flits (left) and 16 flits (right)}
\label{tab:tightness}
\end{table*}
\renewcommand{\arraystretch}{1.5}

To assess the tightness of the delay bounds yielded by G-BATA, we consider herein simulation results using Noxim simulator engine \cite{noxim_catania2016}. 
We have configured Noxim to control the traffic pattern using the provided traffic pattern file option.
For each flow, we have specified:
\begin{itemize}
    \item the source and destination cores;
    \item $pir$, packet injection rate, \ie{} the rate at which packets are sent when the flow is active;
    \item $por$, probability of retransmission, \ie{} the probability one packet will be retransmitted (in our context, this parameter is always 0);
    \item $t_{\rm on}$, the time the flow wakes up, \ie{} starts transmitting packets with the packet injection rate;
    \item $t_{\rm off}$, the time the flow goes to sleep, \ie{} stops transmitting;
    \item $P$, the period of the flow.
\end{itemize}

Moreover, since we want to simulate a deterministic flow behavior to approach the worst-case scenario, we use the following parameters for each flow:
\begin{itemize}
  \item Maximal packet injection rate : 1.0;
  \item Minimal probability of retransmission : 0.0;
\end{itemize}

We also have to pick $t_{\rm on}$, that is determine at what time within its period the flow is going to wake up from its inactivity and start sending packets.
Since we want the flow to be periodic, we set its active period to be as short as possible so that we ensure it wakes up, sends exactly one packet, and goes to sleep until the next period.
To create different contention scenarios and try approaching the worst-case of end-to-end delays, we randomly chose a value of $t_{\rm on}$ for each flow and perform simulations with uniformly distributed values of offsets for each flow.
We generate 40000 different traffic configurations for each set of parameters and simulate each of them for an amount of time allowing at least 5 packets to be transmitted.

We simulate the configuration of Figure \ref{fig:sensitivity_configuration}, when varying buffer sizes in 4, 8 and 16 flits, and flow rates in 8\% and 32\% of the total available bandwidth. 
We run each flow configuration many times while varying the flows offsets. 
We extract the worst-case end-to-end delay found by the simulator over all the simulations, and for each flow $f$, we compute the corresponding ``tightness ratio'' $\tau_f$, that is the ratio of the achievable worst-case delay $D_{\rm WC}$ and the worst-case delay bound $D_f$: 
\[ \tau_f = \frac{D_{\rm WC}}{D_f} \]

We simulate the configuration of Figure \ref{fig:sensitivity_configuration}, when varying buffer sizes in 4, 8 and 16 flits, and flow rates in 8\% and 32\% of the total available bandwidth. 
We extract the worst-case end-to-end delay found by the simulator and compute the tightness ratio for each flow. 
The obtained results are gathered in Table \ref{tab:tightness}. 

We also recall the computed tightness ratios obtained with BATA, detailed in \cite{giroudot2019_tightness_and_computation_assessment}. 
Additionally, we computed and included the congestion indexes associated with G-BATA approach. We only displayed results for buffer sizes 4 and 16. 

We notice that the lower the congestion indexes are, the greater the tightness is.
Low congestion indexes mean that the contending possibilities are reduced.
Hence the worst case is simpler to find and thus more likely to be achieved or approached with randomly chosen offsets.
We stress out the fact that there are many possibilities for the wake up time of each flow, and that our series of simulations may not have been able to approach or achieve the worst-case for every flow. \\

For a buffer size of 4 flits and a flow rate of 8\%, G-BATA and BATA give similar results (with a slightly better average tightness for BATA). 
However, for a 32\% rate, G-BATA gives tighter bounds.
For 16 flits of buffer size, BATA gives tighter results for both rates. 
However, we want to stress out that in this case, for 32\% rate, we might not be able to verify that no CPQ can occur. Thus the results yielded by BATA should be taken with caution. \\

\textbf{Key points:} On the tested configuration, with 4-flit-large buffers and at 8\% flow rate, both models give similar results. 
With the same buffer size and a higher rate (32\%), G-BATA gives tighter results than BATA, showing that BATA tends to be pessimistic for high flow rates. 
With larger buffer sizes, BATA performs better, but when flow rates are high, BATA might not be applicable. 
Overall, the tightness is good. G-BATA averages at 72\% when the buffer size is 4 flits and 56\% for 16 flits, whereas BATA averages at 59\% and 81\%, respectively.
For flows subject to the more complex congestion patterns, the worst-case may not have been approached as closely as for flows undergoing little to no interference, hence the derived tightness ratio is smaller.
This conjecture is supported by the fact that the measured tightness is lower for flows with higher congestion indexes. \\

\subsection{Discussion}
\begin{figure*}[ht]
\centering
\resizebox{0.5\textwidth}{!}{\input{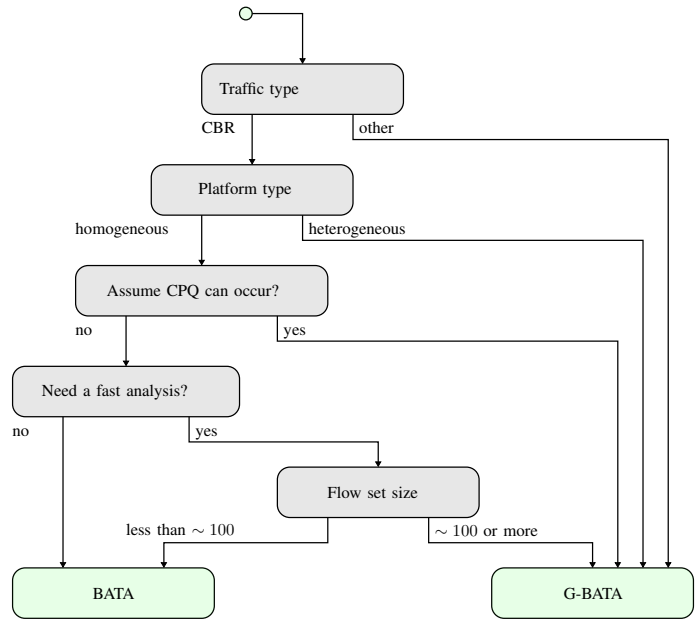}}
\caption{Decision-making graph for the use of BATA \emph{vs} G-BATA}
\label{fig:decision_bata_gbata}
\end{figure*}

In order to determine whether BATA or G-BATA should be used, we propose a decision-making graph (Figure \ref{fig:decision_bata_gbata}). 
The first choices regard the system characteristics. If the traffic is non-CBR, or if the platform is heterogeneous, BATA is not applicable, thus G-BATA should be used. 
With CBR traffic and homogeneous platforms, BATA may be used provided CPQ does not occur, \ie{} provided one packet of a flow cannot catch up on the previously injected one. 

However, due to the computational complexity of BATA for large flowsets, the analysis with BATA may take a long time. Therefore we recommend the use of G-BATA for configurations with more than 80-100 flows. 
The main interest of using BATA when the appropriate assumptions are verified is that it may give tighter results than G-BATA in some cases, \eg{} when buffer size is large compared to packet lengths. 


\section{Automotive Case Study}
\label{sec:case_study}

We now perform our analysis on the case study proposed in \cite{burns2010_schedulability_analysis_real_time_on_chip_communication} and used in \cite{nikolic_rt_analysis_priority_preemptive_nocs}. 
The chosen application is the control of an autonomous vehicle. It features several tasks in charge of processing data from the sensors, managing the obstacle data base, controling the actuators, etc. Various data flows are exchanged between these tasks.

Further description of the application can be found in \cite{burns2010_schedulability_analysis_real_time_on_chip_communication}.
We took the same 33 tasks mapped on a $4 \times 4$ 2D-mesh NoC, and the same mapping of the 38 data flows between tasks, routed in a XY fashion. 

The parameters used are the following: 
\begin{itemize}
    \item The duration of a cycle is 0.5 ns;
    \item All routers have a technological latency of 3 cycles; 
    \item The link capacity is one flit per cycle;
    \item Flows' priority assignment follows a rate monotonic policy; 
    \item Each router supports 4 Virtual Channels with no priority-sharing and no VC-sharing, \ie{}, one flow per VC;
    \item To compare our results to the ones in \cite{nikolic_rt_analysis_priority_preemptive_nocs}, we performed the analysis for different buffer sizes (2, 100 and 1000000 flits, the latest being large enough to assume buffer size is infinite). 
\end{itemize}

\begin{figure*}[!htbp]
\centering
\includegraphics[scale=0.5]{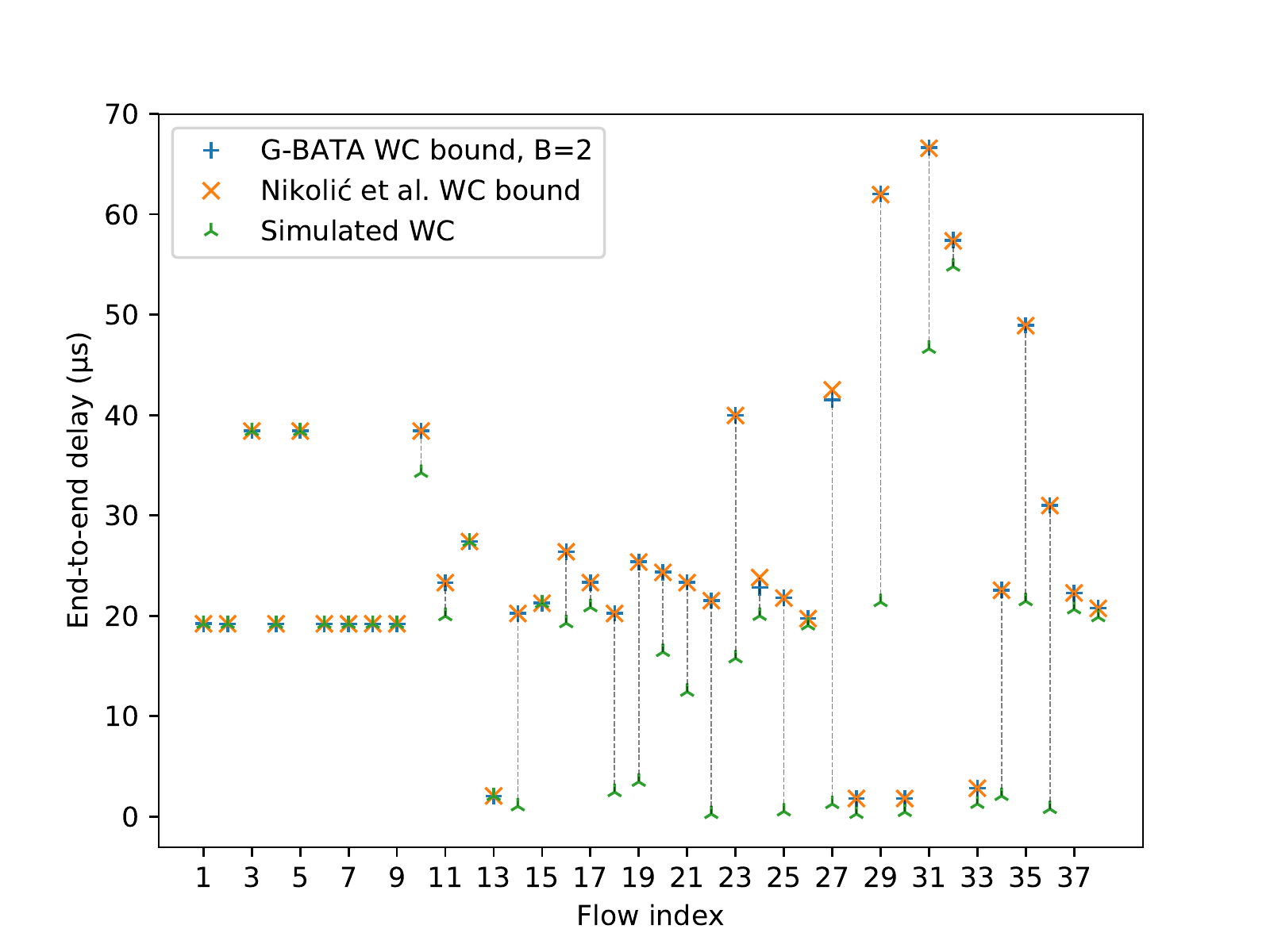}
\includegraphics[scale=0.5]{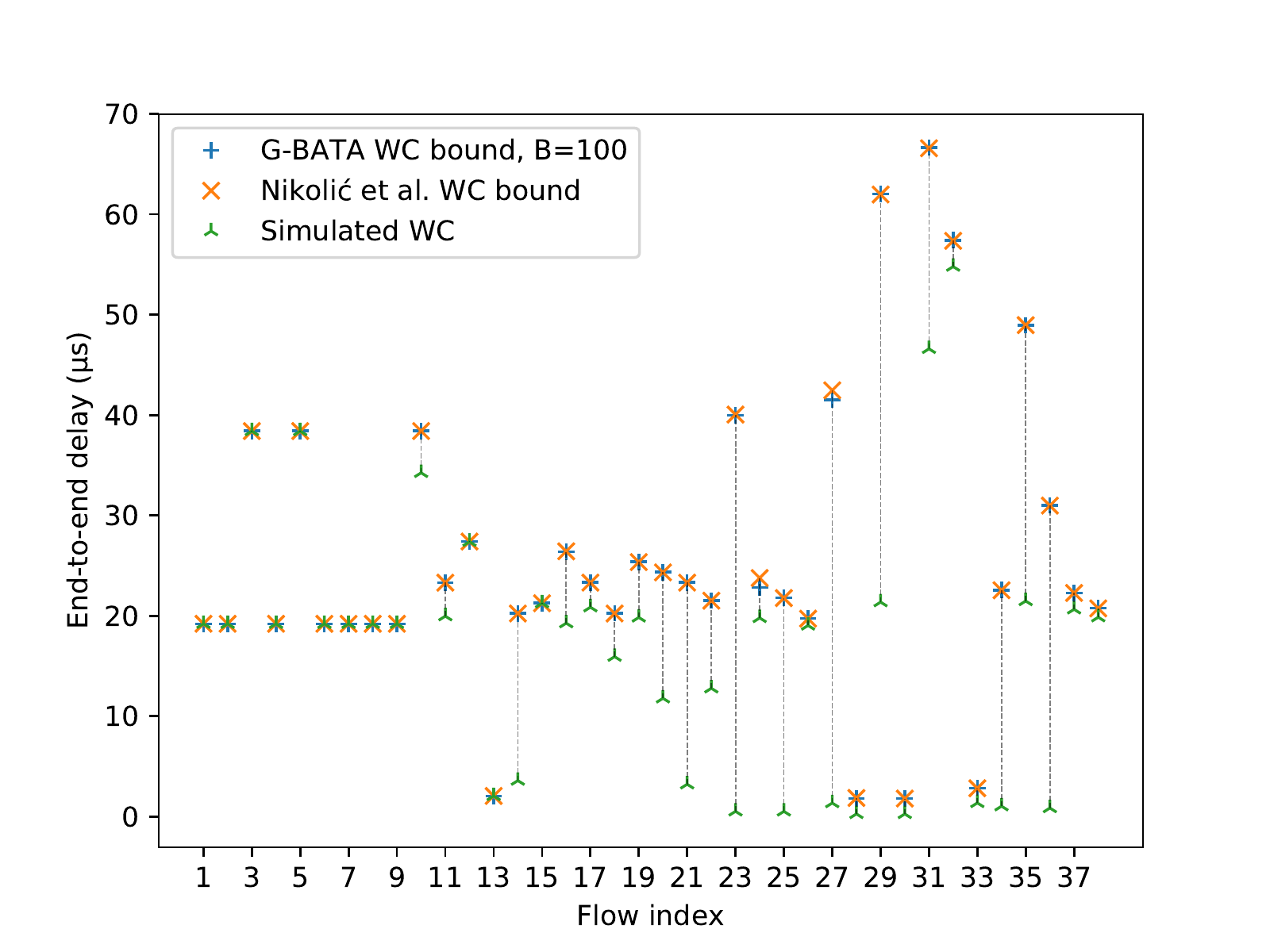}
\includegraphics[scale=0.5]{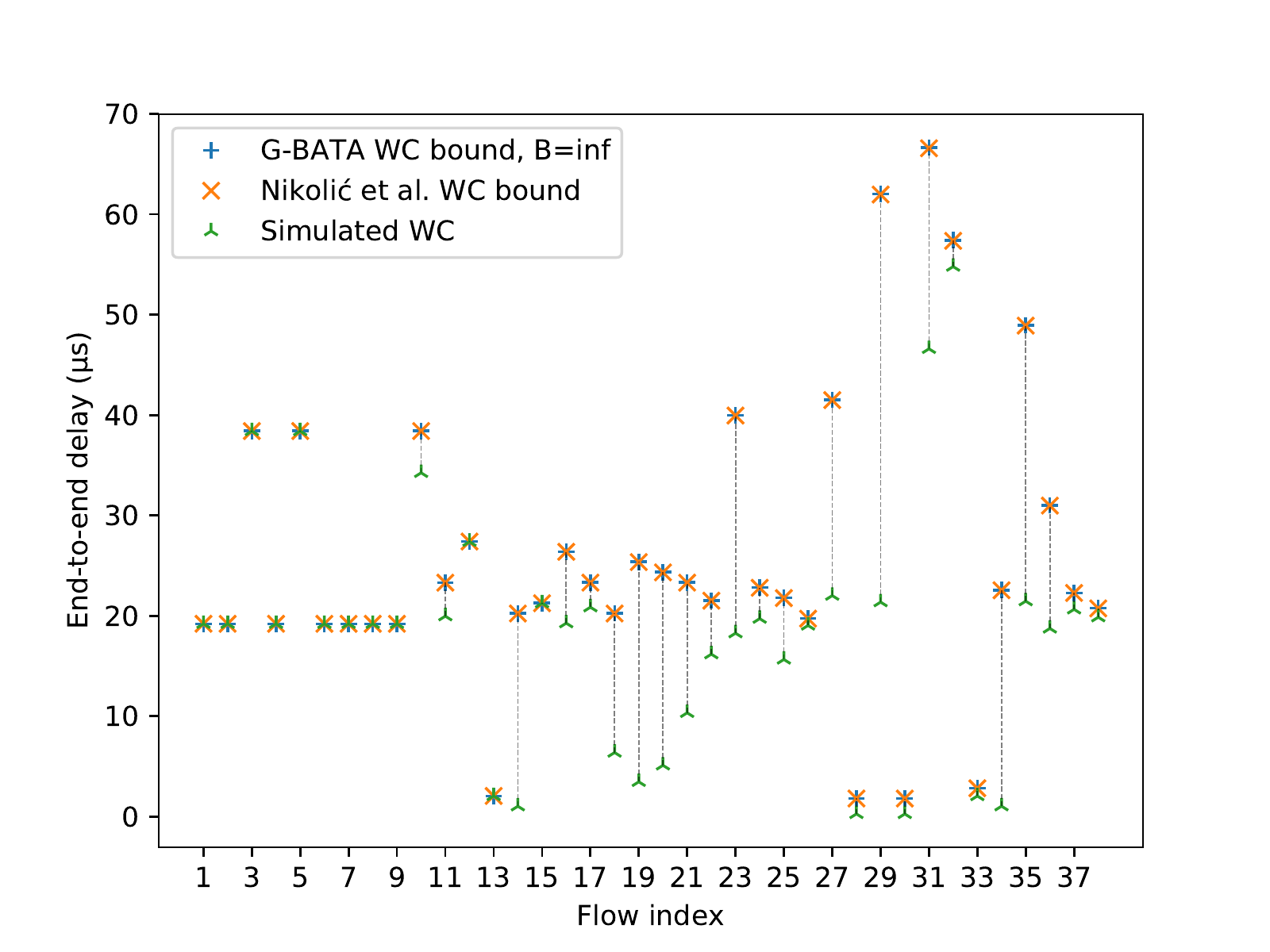}
\caption{Worst-case end-to-end delay bounds comparative}
\label{fig:case_study_results}
\end{figure*}

All flows have a different priority. As they are mapped to VCs in such a way that at each router, all VCs are non-shared, there is no indirect blocking.
Thus, we expect BATA and G-BATA to give the exact same results for the worst-case delay bounds, which we checked was the case.
We then plotted comparative graphs on Figure \ref{fig:case_study_results} and computed the average tightness of our approach (Table \ref{tab:tightness_differences}), using results from simulations performed by Nikolić et al. \cite{nikolic_rt_analysis_priority_preemptive_nocs}. 
The average tightness ratio for G-BATA approach with buffer size 2, 100 and infinite are 64\%, 67\% and 71\% respectively.

We first notice that our approach gives similar results to \cite{nikolic_rt_analysis_priority_preemptive_nocs}. 
To further quantify the similarity of the results, we subtracted the tightness ratio obtained by the two approaches on each bound to obtain what we call ``tightness difference'', denoted $\Delta\tau$. For a given flow:
\[ \Delta\tau = \tau_{\textrm{G-BATA}} - \tau_{\textrm{ST}} \quad, \]
where $\tau_{G-BATA}$ is the tightness ratio of the bound yielded by G-BATA, and $\tau_{ST}$ is the tightness ratio of the bound yielded by the method of \cite{nikolic_rt_analysis_priority_preemptive_nocs}. 
The tightness difference $\Delta\tau$ is positive when G-BATA gives the tighter bound and negative otherwise. 
We synthesized the differences in Table \ref{tab:tightness_differences}. 
We computed the minimum, maximum and average tightness difference.

\renewcommand{\arraystretch}{1.2}
\begin{table}
\centering
\begin{tabular}{@{}lrrr@{}}
\toprule
                              & $B=2$   & $B=100$  & $B=\infty$ \\
\midrule
Average tightness             & 64\%    & 67\%     & 71\%     \\
\midrule
Average tightness difference  & +0.07\% & +0.08\%  & -0.03\%  \\
Maximum tightness difference  & +3.70\% & +3.49\%  & +0.01\%  \\
Minimum tightness difference  & -0.10\% & -0.10\%  & -0.10\%  \\
\bottomrule
\end{tabular}
\medskip
\caption{Average tightness and tightness differences for various buffer sizes, for G-BATA and state-of-the-art approach}
\label{tab:tightness_differences}
\end{table}
\renewcommand{\arraystretch}{1.5}

Even though they are based on fundamentally different theories, we can notice both approaches yield very close results, giving credit to both models. 

Authors in \cite{nikolic_rt_analysis_priority_preemptive_nocs} have shown that 4 VCs are needed to find a mapping of flows to VCs that ensures each flow has exclusive use of the VC within each router, which greatly simplifies the computation.
However, having only one flow per VC at each node can raise scalability problems: with larger and/or less favorable configurations, ensuring each flow has the exclusive use of a VC within each router would require a number of different VCs that is not reasonable any more.

In that respect, we want to stress out that our model allows priority sharing and VC sharing (several flows sharing priority levels and VCs). 
Therefore, we have performed another analysis on the same configuration using only 2 VCs, with the following priority mapping:
\begin{itemize}
    \item Flows 1 to 19 have the higher priority and are mapped to VC0;
    \item Flows 20 to 38 have the lower priority and are mapped to VC1.
\end{itemize}
We also analyze a configuration with only 1 shared VC.

\begin{figure}[!thb]
\centering
\includegraphics[scale=0.5]{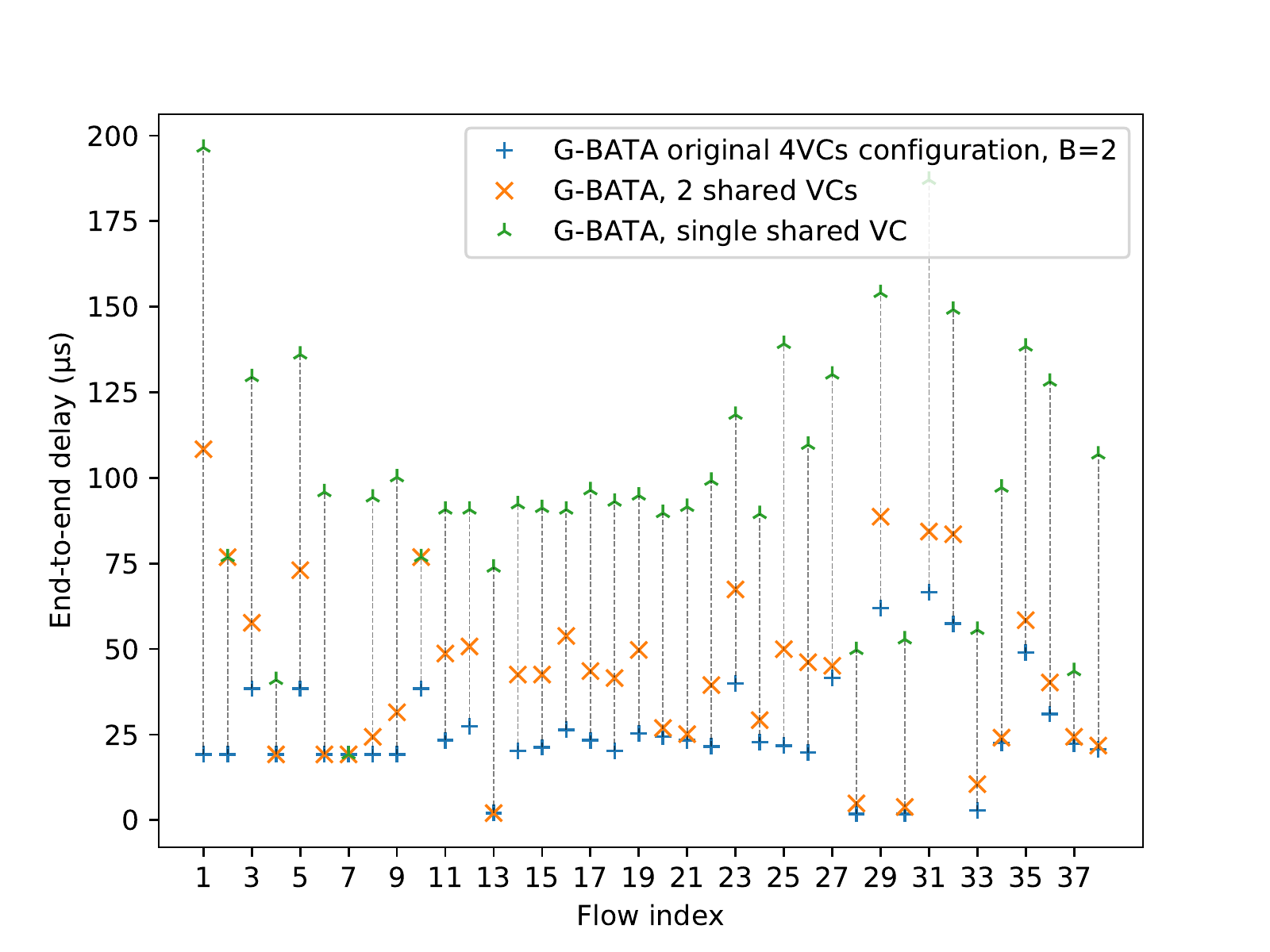}
\caption{Delay bounds with 4, 2 and 1 VC with buffer size = 2 flits}
\label{fig:vc_comparative}
\end{figure}

We have plotted the results with the different VC configurations on Figure \ref{fig:vc_comparative}. We only displayed the results for a buffer size of 2 flits, but the trend is similar with other sizes.
To get an insight into the impact of reducing the number of VCs on delay bounds, we also computed, for each flow and for each $n$ VC configuration, the relative increase of the worst-case delay bounds compared to the delay bound with 4 VCs, as follows:
\begin{eqnarray*}
inc_n &=& \frac{\textrm{delay with $n$ VCs} - \textrm{delay with 4 VCs}}{\textrm{delay with 4VCs}}
\end{eqnarray*}
The results are shown on Table \ref{tab:relative_bound_inc}. 

First, as we can notice from Fig. \ref{fig:vc_comparative}, all flows have delay bounds less than their periods (the shortest period is 40 ms); thus remain schedulable. 
When we reduce the number of VCs, the computed delay bound either increases or remains the same for all flows. 
When mapping all the flows to the same priority level, two factors impact the end-to-end delay bound. 
First, more interference patterns will be possible, especially considering that the configuration with 4 VCs did not allow for any indirect blocking. Hence, additional delay will impact all the flows. 
Second, it is likely that highest priority flows will suffer from additional delay because the arbitration provides equal fairness to all flows. 
Conversely, the lower priority flows will suffer less or not at all from that redistributed fairness, and may even experience lower delays due to competition with other flows. 
In the considered case, however, the additional complexity of the indirect blocking prevails and no flow experiences a decrease of its end-to-end delay bound. 

Moreover, as shown in Table \ref{tab:relative_bound_inc}, the average bound increase stays reasonable (up to few times more than the original one) when the number of available VCs is divided by up to 4. Hence, G-BATA yields noticeable improvements to decrease the platform complexity (less Virtual Channels) while guaranteeing schedulability, in comparison to the state-of the-art method in \cite{nikolic_rt_analysis_priority_preemptive_nocs}.
\renewcommand{\arraystretch}{1.2}
\begin{table}
\centering
\begin{tabular}{@{}lrr@{}}
\toprule
G-BATA                           & 2 VCs    & 1 VC      \\
\midrule                         
Average bound increase           & 81.62\%  & 548.33\%  \\ 
Minimal bound increase           & 0.00\%   & 0.00\%    \\ 
Maximal bound increase           & 464.16\% & 3490.30\% \\
\bottomrule
\end{tabular}
\medskip
\caption{Relative increase of the worst-case end-to-end delay bounds for $B=2$, for 2 VC and one VC configurations \emph{vs} the 4 VC configuration, for G-BATA appproach}
\label{tab:relative_bound_inc}
\end{table}
\renewcommand{\arraystretch}{1.5}

Finally, we provide some insights into the runtime of G-BATA under different VC-configurations. For each buffer size and number of VCs, we measured
the runtime of G-BATA and summarized our results in Table \ref{tab:case_study_runtime}. We notice runtimes with non-shared VCs are in the order of 10 to 100 times lower than runtimes
with 1 and 2 VCs. This confirms our conclusions regarding the inherent complexity of G-BATA, shown in Section \ref{subsec:computational_gbata}, when enabling the priority-sharing and VC-sharing assumptions.

When no VC is shared between several flows, the IB latency is zero, and consequently, computing the end-to-end service curve is faster. 
On the contrary, when VCs are shared, there are \begin{enumerate*}[label=(\roman*)] \item additional recursive calls to end-to-end service curve function needed to compute the IB latency; 
and \item a more complex interference graph to construct. 
\end{enumerate*}
Therefore, we can expect an increase in the analysis duration. 
In the studied cases, G-BATA still performs in a very reasonable duration (one second or less). 

\renewcommand{\arraystretch}{1}
\begin{table}
\centering
\begin{tabular}{@{}llrrr@{}}
\toprule
                      &            & 4 VCs     & 2 VCs          & 1 VC    \\
\midrule
Runtime of G-BATA (ms) & $B=2$      & 4.77      & 62.2           & 316     \\
                      & $B=100$    & 4.95      & 55.8           & 306     \\
                      & $B=\infty$ & 4.78      & 217            & 1346    \\
\bottomrule
\end{tabular}
\medskip
\caption{Runtimes of G-BATA for different NoC configurations}
\label{tab:case_study_runtime}
\end{table}
\renewcommand{\arraystretch}{1.5}

\section{Conclusions and Perspectives}
\label{sec:conclusion}

Starting from the observation that bursty traffic is not covered by our previously published BATA model, we aimed at extending the proposed analysis to handle CPQ scenarios, thus including bursty traffic flows, and heterogeneous architectures. 
First, we extended the notions developped in \cite{giroudot_buffer_aware} to model heterogeneous architectures
and exhibited a CPQ scenario to explain the idea of our extension.

Then, we proposed a new approach, G-BATA, improving the indirect blocking analysis based on dependency graphs to capture interference patterns involving CPQ. 
Following this, we adapted the indirect blocking latency computation to take into account the new way of modeling indirect blocking patterns, and consequently decreased the number of recursive calls needed to compute end-to-end service curves.

Finally, we evaluated our approach on several aspects: \begin{enumerate*}[label=(\roman*)] \item we studied the sensitivity of the model to input parameters such as router buffer size, flow rate, and packet length. 
We found that increasing buffer size does not reduce the end-to-end delay bound, and that for a given flow rate, sending small packets is more worst-case-efficient than sending big packets. We also found that BATA bounds are generally more pessimistic than G-BATA when flow rates increase; 
\item we evaluated the scalability of our approaches. We showed BATA hardly scales beyond configurations of 50 to 100 flows, whereas G-BATA is able to analyze 800-flow configurations in a reasonable time (below 10 seconds per flow); 
\item we evlauated the tightness of the model given bound on a test-case and achieved an average tightness ratio of 71\%, without; 
\item we confronted our model to a realistic case-study to further check for the correctness of the bounds and the efficiency of G-BATA in comparison to a state-of-the-art approach.
\end{enumerate*}

In a future work, we plan to focus on addressing related problems such as Software/Hardware mapping. 
We would like to include our dependency graph-based approach in Design Space Exploration techniques of manycore platforms. 
This would also allow us to tackle more complex case studies and improve the system performance.

\bibliography{buffer_aware_noc_analysis}{}
\bibliographystyle{ieeetr}

\end{document}